%% file: main.tex
\documentclass[11pt]{article}

\usepackage[utf8]{inputenc} 
\usepackage[T1]{fontenc}    
\usepackage{hyperref}       
\usepackage{url}            
\usepackage{booktabs}       
\usepackage{amsfonts}       
\usepackage{nicefrac}       
\usepackage{microtype}      
\usepackage{lipsum}
\usepackage{graphicx}
\usepackage{comment}
\usepackage{outlines}
\usepackage{enumitem}
\graphicspath{{media/}}
\usepackage{tabularx}
\usepackage{todo}
\usepackage{colortbl}
\usepackage{booktabs}
\usepackage{array}
\usepackage{float}
\usepackage[table]{xcolor}
\usepackage{caption} 
\usepackage{newtxtext,newtxmath}
\usepackage{pdflscape}
\usepackage{threeparttable}
\usepackage{authblk, tikz}
\usepackage{dcolumn, ulem}
\usepackage{fancyhdr}
\usepackage{subfigure}
\usepackage[round]{natbib}
\usepackage[margin=1in]{geometry}

\title{GPTs are GPTs: An Early Look at the Labor Market Impact Potential of Large Language Models} 

\author[1]{Tyna Eloundou}
\author[1,2]{Sam Manning}
\author[1]{Pamela Mishkin\thanks{Corresponding author (pamela@openai.com). Authors contributed equally and are listed alphabetically.}}
\author[3]{Daniel Rock}
\affil[1]{OpenAI}
\affil[2]{OpenResearch}
\affil[3]{University of Pennsylvania}
\fancypagestyle{plain}{
  \fancyhf{}
  \fancyhead[C]{WORKING PAPER}
}

\begin{document}
\fancyhf{}
\fancyhead[C]{WORKING PAPER}
\maketitle

\begin{abstract}

We investigate the potential implications of large language models (LLMs), such as Generative Pre-trained Transformers (GPTs), on the U.S. labor market, focusing on the increased capabilities arising from LLM-powered software compared to LLMs on their own. Using a new rubric, we assess occupations based on their alignment with LLM capabilities, integrating both human expertise and GPT-4 classifications. Our findings reveal that around 80\% of the U.S. workforce could have at least 10\% of their work tasks affected by the introduction of LLMs, while approximately 19\% of workers may see at least 50\% of their tasks impacted. We do not make predictions about the development or adoption timeline of such LLMs. The projected effects span all wage levels, with higher-income jobs potentially facing greater exposure to LLM capabilities and LLM-powered software. Significantly, these impacts are not restricted to industries with higher recent productivity growth. Our analysis suggests that, with access to an LLM, about 15\% of all worker tasks in the US could be completed significantly faster at the same level of quality. When incorporating software and tooling built on top of LLMs, this share increases to between 47 and 56\% of all tasks. This finding implies that LLM-powered software will have a substantial effect on scaling the economic impacts of the underlying models. We conclude that LLMs such as GPTs exhibit traits of general-purpose technologies, indicating that they could have considerable economic, social, and policy implications.

\end{abstract}

\section{Introduction}

\begin{figure}
    \centering
    \includegraphics[width=.9\textwidth]{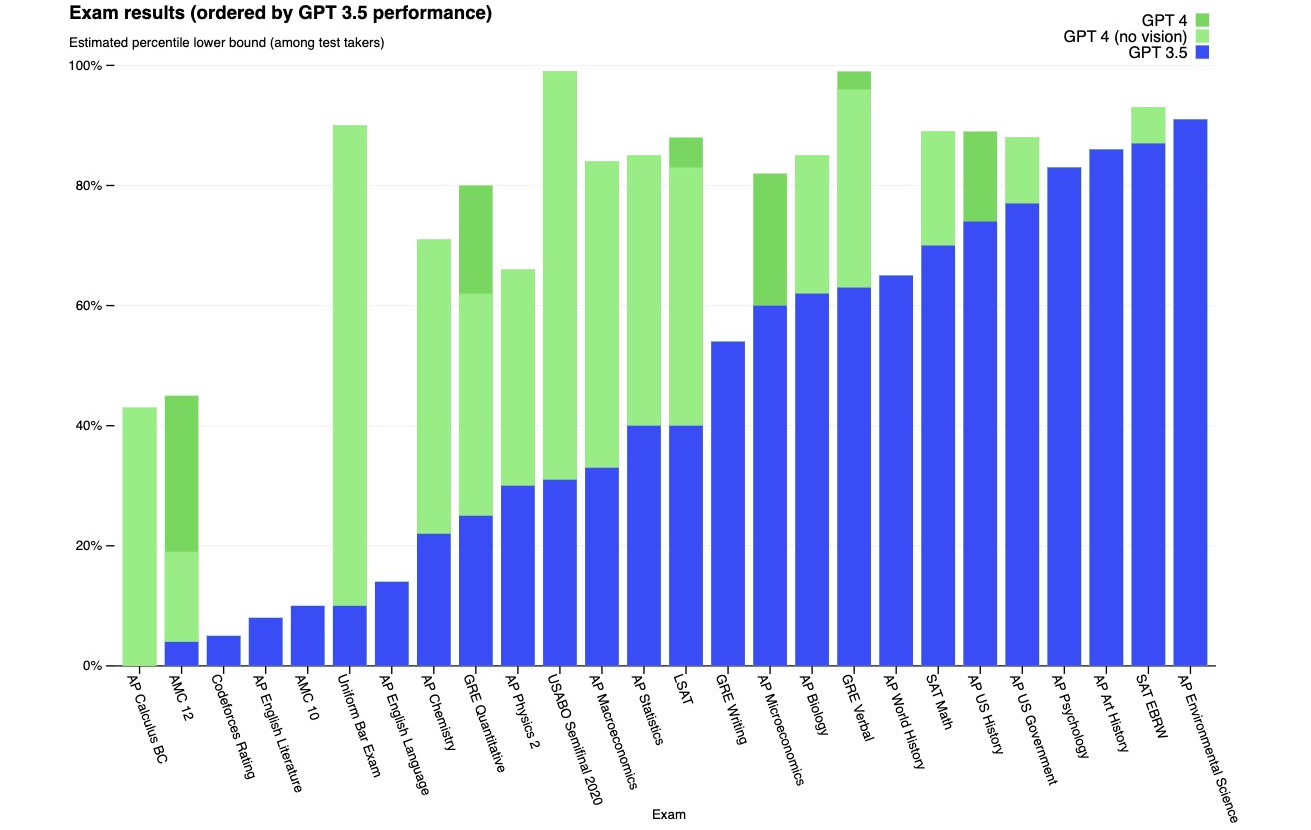}
    \caption{Taken directly from GPT-4 Technical Report \citep{gpt4}. To get a sense of how quickly model capabilities are progressing -- consider the jump in exam performance between GPT-3.5 and GPT-4 \citep{gpt4}.}
    \label{fig:exames}
\end{figure}

As shown in Figure \ref{fig:exames}, recent years, months, and weeks have seen remarkable progress in the field of generative AI and large language models (LLMs). While the public often associates LLMs with various iterations of the Generative Pre-trained Transformer (GPT), LLMs can be trained using a range of architectures, and are not limited to transformer-based models \citep{47751}. LLMs can process and produce various forms of sequential data, including assembly language, protein sequences and chess games, extending beyond natural language applications alone. In this paper, we use LLMs and GPTs somewhat interchangeably, and specify in our rubric that these should be considered similar to the GPT-family of models available via ChatGPT or the OpenAI Playground (which at the time of labeling included models in the GPT-3.5 family but not in the GPT-4 family). We examine LLMs with text- and code-generating abilities, use the term "generative AI" to additionally include modalities such as images or audio, and use "LLM-powered software" to cover tools built on top of LLMs or that combine LLMs with other generative AI models.

Our study is motivated less by the progress of these models alone though, and more by the breadth, scale, and capabilities we've seen in the complementary technologies developed around them. The role of complementary technologies remains to be seen, but maximizing the impact of LLMs appears contingent on integrating them with larger systems \citep{bresnahan2019artificial, agrawal2021ai}. While the focus of our discussion is primarily on the generative capabilities of LLMs, it is important to note that these models can also be utilized for various tasks beyond text generation. For example, embeddings from LLMs can be used for custom search applications, and LLMs can perform tasks such as summarization and classification where the context may be largely contained in the prompt.

To complement predictions of technology's impacts on work and provide a framework for understanding the evolving landscape of language models and their associated technologies, we propose a new rubric for assessing LLM capabilities and their potential effects on jobs. This rubric (\ref{exposure_tax}) measures the overall exposure of tasks to LLMs, following the spirit of prior work on quantifying exposure to machine learning \citep{Brynjolfsson2018, SeamansRajFelten2018, Webb2020}. We define exposure as a proxy for potential economic impact without distinguishing between labor-augmenting or labor-displacing effects. We employ human annotators and GPT-4 itself as a classifier to apply this rubric to occupational data in the U.S. economy, primarily sourced from the O*NET database.\footnote{This is distinct from recent social science research that makes use of LLMs to simulate human behavior \citep{horton2023large, Sorensen_2022}}\footnote{While our exposure rubric does not necessarily tie the concept of language models to any particular model, we were strongly motivated by our observed capabilities of GPT-4 and the suite of capabilities we saw in development with OpenAI's launch partners \citep{gpt4}.}

To construct our primary exposure dataset, we collected both human annotations and GPT-4 classifications, using a prompt tuned for agreement with a sample of labels from the authors. We observe similar agreement levels in GPT-4 responses and between human and machine evaluations, when aggregated to the task level. This exposure measure reflects an estimate of the technical capacity to make human labor more efficient; however, social, economic, regulatory, and other determinants imply that technical feasibility does not guarantee labor productivity or automation outcomes. Our analysis indicates that approximately 19\% of jobs have at least 50\% of their tasks exposed when considering both current model capabilities and anticipated tools built upon them. Human assessments suggest that only 3\% of U.S. workers have over half of their tasks exposed to LLMs when considering existing language and code capabilities without additional software or modalities. Accounting for other generative models and complementary technologies, our human estimates indicate that up to 49\% of workers could have half or more of their tasks exposed to LLMs. 

Our findings consistently show across both human and GPT-4 annotations that most occupations exhibit some degree of exposure to LLMs, with varying exposure levels across different types of work. Occupations with higher wages generally present with higher exposure, a result contrary to similar evaluations of overall exposure to machine learning \citep{brynjolfssonQuantifyingDistributionMachine2023}. When regressing exposure measures on skillsets using O*NET's skill rubric, we discover that roles heavily reliant on science and critical thinking skills show a negative correlation with exposure, while programming and writing skills are positively associated with LLM exposure. Following \citet{autor2022new}, we examine barriers to entry by "Job Zones" and find that occupational exposure to LLMs weakly increases with the difficulty of job preparation. In other words, workers facing higher (lower) barriers to entry in their jobs tend to experience more (less) exposure to LLMs.

We further compare our measurements to previous efforts documenting the distribution of automation exposure in the economy and find broadly consistent results. Most other technology exposure measures we examine are statistically significantly correlated with our preferred exposure measure, while measures of manual routineness and robotics exposure show negative correlations. The variance explained by these earlier efforts \citep{acemoglu2011skills, FreyOsborne2017, Brynjolfsson2018, SeamansRajFelten2018, Webb2020, brynjolfssonQuantifyingDistributionMachine2023}, along with wage controls, ranges from 60 to 72\%, indicating that 28 to 40\% of the variation in our AI exposure measure remains unaccounted for by previous technology exposure measurements. 

We analyze exposure by industry and discover that information processing industries (4-digit NAICS) exhibit high exposure, while manufacturing, agriculture, and mining demonstrate lower exposure. The connection between productivity growth in the past decade and overall LLM exposure appears weak, suggesting a potential optimistic case that future productivity gains from LLMs may not exacerbate possible cost disease effects \citep{baumol2012cost, aghion2018artificial}. \footnote{Baumol's cost disease is a theory that explains why the cost of labor-intensive services, such as healthcare and education, increases over time. This happens because wages for skilled workers in other industries increase, but there is no corresponding increase in productivity or efficiency in these service industries. Therefore, the cost of labor in these industries becomes relatively more expensive compared to other goods and services in the economy.} 

Our analysis indicates that the impacts of LLMs like GPT-4, are likely to be pervasive. While LLMs have consistently improved in capabilities over time, their growing economic effect is expected to persist and increase even if we halt the development of new capabilities today. We also find that the potential impact of LLMs expands significantly when we take into account the development of complementary technologies. Collectively, these characteristics imply that Generative Pre-trained Transformers (GPTs) are general-purpose technologies (GPTs).\footnote{For the remainder of the paper we spell out general-purpose technologies when it is used outside of stating "GPTs are GPTs."} \citep{bresnahan1995general, lipsey2005economic}. 

\citep{goldfarb2023could} argue that machine learning as a broad category is likely a general-purpose technology. Our evidence supports a wider impact, as even subsets of machine learning software meet the criteria for general-purpose technology status independently. This paper's primary contributions are to provide a set of measurements of LLM impact potential and to demonstrate the use case of applying LLMs to develop such measurements efficiently and at scale. Additionally, we showcase the general-purpose potential of LLMs. If "GPTs are GPTs," the eventual trajectory of LLM development and application may be challenging for policymakers to predict and regulate. As with other general-purpose technologies, much of these algorithms' potential will emerge across a broad range of economically valuable use cases, including the creation of new types of work \citep{acemoglu2018race, autor2022new}. Our research serves to measure what is technically feasible now, but necessarily will miss the evolving impact potential of the LLMs over time.

The paper is structured as follows: Section \ref{sec:litreview} reviews relevant prior work, Section \ref{sec:data} discusses methods and data collection, Section \ref{sec:econimpact} presents summary statistics and results, Section \ref{sec:validation} relates our measurements to earlier efforts, Section \ref{sec:discussion} discusses the results, and Section \ref{sec:conclusion} offers concluding remarks.

\section{Literature Review}
\label{sec:litreview}

\subsection{The Advancement of Large Language Models}
\label{sec:llmlit}
In recent years, generative AI models have gained significant attention from both the artificial intelligence (AI) research community and the general public, due to their ability to tackle a wide array of complex language-based tasks. The progress in these models' abilities has been fueled by multiple factors, including increased model parameter count, greater training data volume, and enhanced training configurations \citep{brown2020language, radford2019language, hernandez2021scaling, kaplan2020scaling}. Broad, state-of-the-art LLMs, such as LaMDA \citep{thoppilan2022lamda} and GPT-4 \citep{gpt4}, excel in diverse applications like translation, classification, creative writing, and code generation—capabilities that previously demanded specialized, task-specific models developed by expert engineers using domain-specific data.

Concurrently, researchers have improved the steerability, reliability, and utility of these models using methods like fine-tuning and reinforcement learning with human feedback \citep{ouyang2022training, bai_training_2022}. These advancements enhance the models' ability to discern user intent, rendering them more user-friendly and practical. Moreover, recent studies reveal the potential of LLMs to program and control other digital tools, such as APIs, search engines, and even other generative AI systems \citep{schick2023toolformer, mialon2023augmented, langchain}. This enables seamless integration of individual components for better utility, performance, and generalization. At their limit, these trends suggest a world where LLMs may be capable of executing any task typically performed at a computer.

Generative AI models have mostly been deployed as modular specialists, performing specific tasks such as generating images from captions or transcribing text from speech. However, we argue that it is essential to view LLMs as versatile building blocks for creating additional tools. Developing these tools and integrating them into systems will require time and possibly significant reconfiguration of existing processes across various industries. Nevertheless, we are already witnessing emerging adoption trends. Despite their limitations, LLMs are increasingly being integrated into specialized applications in fields like writing assistance, coding, and legal research. These specialized applications then allow businesses and individuals to adopt LLMs into their workflows.

We emphasize the significance of these complementary technologies, partly because out-of-the-box general-purpose LLMs may continue to be unreliable for various tasks due to issues such as factual inaccuracies, inherent biases, privacy concerns, and disinformation risks \citep{abid_persistent_2021, schramowski_large_2022, goldstein_generative_2023, 4systemcard}. However, specialized workflows—including tooling, software, or human-in-the-loop systems—can help address these shortcomings by incorporating domain-specific expertise. For example, \href{https://casetext.com/}{Casetext} offers LLM-based legal research tools that provide lawyers with quicker and more accurate legal research results, utilizing embeddings and summarization to counter the risk that GPT-4 could provide inaccurate details about a legal case or set of documents. \href{https://github.com/features/copilot}{GitHub Copilot} is a coding assistant that employs LLMs to generate code snippets and auto-complete code, which users can then accept or reject based on their expertise. In other words, while it's true that on its own GPT-4 does not "know what time it is," it's easy enough to give it a watch.

Furthermore, a positive feedback loop may emerge as LLMs surpass a specific performance threshold, allowing them to assist in building the very tooling that enhances their usefulness and usability across various contexts. This could lower the cost and engineering expertise required to create such tools, potentially accelerating LLM adoption and integration even further \citep{chen2021evaluating, peng2023impact}. LLMs can also become valuable assets in machine learning model development—serving as coding assistants for researchers, data labeling services, or synthetic data generators. There is potential for such models to contribute to economic decision-making at the task level, for instance, by refining methods for task and sub-task allocation between humans and machines \citep{Singla2015LearningTH,Shahaf2010GeneralizedTM}. As LLMs advance over time and better align with user preferences, we can anticipate continuous improvement in performance. However, it is essential to recognize that these trends also bring a variety of serious risks. \citep{hazard_analysis, Weidinger2022, solaiman_release}

\subsection{The Economic Impacts of Automation Technologies}

A large and growing body of literature addresses the labor market impacts of AI and automation technologies. The concept of skill-biased technological change and the task model of automation—often considered the standard framework for understanding technology's influence on labor—originated from research demonstrating that technological progress raises the demand for skilled workers over unskilled workers \citep{katz1992changes}. Numerous studies have built upon this concept, exploring the effects of technological change and automation on workers within a task-based framework \citep{autor2003skill, acemoglu_autor_2011, acemoglu2018race}. This strand of research has shown that workers involved in routine and repetitive tasks are at a higher risk of technology-driven displacement, a phenomenon known as routine-biased technological change. More recent studies have distinguished between technology's task-displacement and task-reinstatement effects (where new technology increases the need for a wider array of labor-intensive tasks) \citep{acemoglu2018race, acemoglu2019automation}. Several studies have shown that automation technologies have resulted in wage inequality in the US, driven by relative wage declines for workers specializing in routine tasks \citep{autor2006polarization, van2011wage, acemoglu2022tasks}. 

Prior research has employed various approaches to estimate the overlap between AI capabilities and the tasks and activities workers undertake in different occupations. These methods include mapping patent descriptions to worker task descriptions \citep{Webb2020, meindl2021exposure}, linking AI capabilities to occupational abilities documented in the O*NET database \citep{SeamansRajFelten2018, felten2023will}, aligning AI task benchmark evaluations with worker tasks via cognitive abilities \citep{Tolan2021}, labeling automation potential for a subset of US occupations and using machine learning classifiers to estimate this potential for all other US occupations \citep{FreyOsborne2017}, modeling task-level automation and aggregating the results to occupation-level insights \citep{arntz2017revisiting}, collecting expert forecasts \citep{grace2018will}, and most relevantly to this paper, devising a new rubric to assess worker activities for their suitability for machine learning \citep{Brynjolfsson2018, brynjolfssonQuantifyingDistributionMachine2023}. Some of these approaches have found exposure to AI technologies at the task-level tends to be diversified within occupation. Considering each job as a bundle of tasks, it would be rare to find any occupation for which AI tools could do nearly all of the work. \citep{autor2022new} finds as well that automation and augmentation exposures tend to be positively correlated. There is also a growing set of studies examining specific economic impacts and opportunities for LLMs \citep{bommasani2021opportunities, felten2023will, korinek2023language, mollick2022new, noy2023experimental, peng2023impact}. Alongside this work, our measurements help characterize the broader potential relevance of language models to the labor market.

General-purpose technologies (e.g. printing, the steam engine) are characterized by widespread proliferation, continuous improvement, and the generation of complementary innovations \citep{bresnahan1995general, lipsey2005economic}. Their far-reaching consequences, which unfold over decades, are difficult to anticipate, particularly in relation to labor demand \citep{bessen2018artificial, korinek2018artificial, acemoglu2020ai, benzell2021}. The realization of general purpose technologies' full potential requires extensive co-invention \citep{bresnahan1995general, bresnahan1996technical, bresnahan2002information,lipsey2005economic, dixon2021robot}, a costly and time-consuming process involving the discovery of new business procedures \citep{david1990dynamo,bresnahan1999computerisation, frey2019technology,brynjolfsson2021productivity,feigenbaum2021organizational}. Consequently, many studies of machine learning technologies focus on systems-level adoption, arguing that organizational systems may require redesign to effectively take advantage of novel machine learning advancements \citep{bresnahan2019artificial, agrawal2021ai, goldfarb2023could}. Appropriately designed systems can yield considerable business value and improve firm performance \citep{rock2019engineering, babina2021artificial, zolas2021advanced}, with AI tools facilitating the discovery process \citep{cockburn2018impact, cheng2022innovae}. By employing task-level information to assess whether LLMs fulfill the criteria of a general purpose technology, we seek to merge the two perspectives for understanding the technology-labor relationship.

We attempt to build on these diverse literature streams in several ways. Echoing \citep{felten2023will}, we focus our analysis on the impact of LLMs, rather than addressing machine learning or automation technologies more broadly. Additionally, we propose a novel method that employs LLMs, specifically GPT-4, to assess tasks for exposure and automation potential, thereby bolstering human scoring efforts. Subsequently, we aggregate our findings to occupations and industries, capturing the overall potential exposure in the contemporary U.S. labor market.

\section{Methods and Data Collection}
\label{sec:data}

\subsection{Data on Activities and Tasks Performed by Occupation in the US}
\label{subsec:task_data}

We use the O*NET 27.2 database \citep{onet272}, which contains information on 1,016 occupations, including their respective Detailed Work Activities (DWAs) and tasks. A DWA is a comprehensive action that is part of completing task, such as "Study scripts to determine project requirements." A task, on the other hand, is an occupation-specific unit of work that may be associated with zero, one, or multiple DWAs. We offer a sample of tasks and DWAs in Table \ref{tab:onet}. The two datasets we use consist of:
\begin{itemize}
    \item 19,265 tasks, consisting of a "task description" and a corresponding occupation, and
    \item 2,087 DWAs, where most DWAs are connected to one or more tasks, and tasks may be associated with one or more DWAs, though some tasks lack any associated DWAs.
\end{itemize}

\input{tables/onet_tables}

\subsection{Data on Wages, Employment, and Demographics}

We obtain employment and wage data from the 2020 and 2021 Occupational Employment series provided by the Bureau of Labor Statistics. This dataset encompasses occupational titles, the number of workers in each occupation, and occupation-level employment projections for 2031, typical education required for entry in an occupation and on-the-job training required to attain competency in an occupation \citep{bls_employment_occupation}. We use the BLS-recommended crosswalk to O*NET \citep{bls_crosswalk} to link the O*NET task and DWA dataset and the BLS Labor Force Demographics \citep{bls_demographics}, which is derived from the Current Population Survey (CPS). Both of these data sources are collected by the U.S. government and primarily capture workers who are not self-employed, are documented, and are working in the so-called formal economy.

\subsection{Exposure}

We present our results based on an exposure rubric, in which we define \textbf{exposure} as a measure of whether access to an LLM or LLM-powered system would reduce the time required for a human to perform a specific DWA or complete a task by at least 50 percent. Though GPT-4 has vision capabilities \cite{gpt4} and "LLM" is often used to refer to a much wider range of modalities, vision and image capabilities were only included in our definition of LLM-powered software. We provide a summary of our rubric below, while the complete rubric can be found in \ref{exposure_tax}. When we have labels for DWAs, we first aggregate them to the task level before aggregating to the occupation level.

\vspace{20pt}

\tikzstyle{mybox} = [draw=black, fill=black!2, very thick,
    rectangle, rounded corners, inner sep=20pt, inner ysep=20pt]
\tikzstyle{fancytitle} =[fill=black, text=white, rounded corners]

\begin{tikzpicture}\small
\node [mybox] (box){%
    \begin{minipage}{0.8\textwidth}
  No exposure (E0) if:
    \begin{itemize}[itemsep=1pt,parsep=1pt,topsep=1pt] 
        \item using the described LLM results in no or minimal reduction in the time required to complete the activity or task while maintaining equivalent quality\footnote{Equivalent quality means that a third party, typically the recipient of the output, would not notice or care about LLM assistance.} or
        \item using the described LLM results in a decrease in the quality of the activity/task output.
    \end{itemize}
     Direct exposure (E1) if:
    \begin{itemize}[itemsep=1pt,parsep=1pt,topsep=1pt] 
        \item using the described LLM via ChatGPT or the OpenAI playground can decrease the time required to complete the DWA or task by at least half (50\%).
    \end{itemize}
     LLM+ Exposed (E2) if:
    \begin{itemize}[itemsep=1pt,parsep=1pt,topsep=1pt] 
        \item access to the described LLM alone would not reduce the time required to complete the activity/task by at least half, but
        \item additional software could be developed on top of the LLM that could reduce the time it takes to complete the specific activity/task with quality by at least half. Among these systems, we count access to image generation systems.\footnote{In practice, as can be seen in the full rubric in Appendix \ref{exposure_tax}, we categorize access to image capabilities separately (E3) to facilitate annotation, though we combine E2 and E3 for all analyses.}
\end{itemize}
    \end{minipage}
};
\node[fancytitle, right=10pt] at (box.north west) {Summary of exposure rubric};
\end{tikzpicture}

\vspace{20pt}

We set the exposure threshold at a potential 50\% reduction in time required to complete a specific DWA or task while maintaining consistent quality. We anticipate that adoption will be highest and most immediate for applications that realize a considerable increase in productivity. Although this threshold is somewhat arbitrary, it was selected for ease of interpretation by annotators. Moreover, regardless of the chosen threshold, we guessed that the real-world reduction in task time would likely be slightly or significantly lower than our estimates, leading us to opt for a relatively high threshold. In our own validation labeling, we found that this corresponded closely to whether an LLM or LLM-powered software could perform the core part of a task or nearly the entire task.

\begin{table}[h!]
\centering
\begin{tabular}{@{}lllrr@{}}
\toprule
\textbf{Comparison} & \multicolumn{1}{c}{$\mathbf{\gamma}$}& \multicolumn{1}{c}{\textbf{Weighting}} & \multicolumn{1}{c}{\textbf{Agreement}} & \multicolumn{1}{c}{\textbf{Pearson's}} \\ \midrule

GPT-4, Rubric 1; Human & $\alpha$ & E1 & 80.8\% & 0.223 \\
                                & $\beta$ & E1 + .5*E2 & 65.6\% & 0.591 \\
                                & $\zeta$ & E1 + E2 & 82.1\% & 0.654 \\ \midrule
GPT-4, Rubric 2; Human & $\alpha$ & E1 & 81.8\% & 0.221 \\
                                & $\beta$ & E1 + .5*E2 & 65.6\% & 0.538 \\
                                & $\zeta$ & E1 + E2 & 79.5\% & 0.589 \\ \midrule
GPT-4, Rubric 1; GPT-4, Rubric 2 & $\alpha$ & E1 & 91.1\% & 0.611 \\
                                  & $\beta$ & E1 + .5*E2 & 76.0\% & 0.705 \\
                                  & $\zeta$ & E1 + E2 & 82.4\% & 0.680 \\ \bottomrule
\end{tabular}
\caption{Model and human comparison of agreement and Pearson's correlation scores. The agreement score is determined by looking at how often the two groups agree on the annotation (e.g. E0, E1 or E2). In the paper we use GPT-4, Rubric 1. Core tasks are given twice the weight at the occupation-level as supplemental tasks. All weights sum to one.}
\label{tab:comparison}
\end{table}

We then collected both human and GPT-4-generated annotations using the exposure rubric, which underlie the bulk of the analyses in this paper.
\begin{itemize}
    \item \textit{Human Ratings:} We obtained human annotations by applying the rubric to each O*NET Detailed Worker Activity (DWA) and a subset of all O*NET tasks and then aggregated those DWA and task scores\footnote{The authors annotated DWAs that clearly required a high degree of physicality or manual dexterity, and the contracted annotators labeled the remaining activities, along with a subset of tasks including those without associated DWAs and those for which there was no clear task-level annotation after aggregating the DWA annotations.} at the task and occupation levels. The authors personally labeled a large sample of tasks and DWAs and enlisted experienced human annotators who have reviewed GPT-3, GPT-3.5 and GPT-4 outputs as part of OpenAI's alignment work \citep{ouyang2022training}.
    \item \textit{GPT-4 Ratings:} We administered a similar rubric to an early version of GPT-4 \citep{gpt4} but on all task/occupation pairs rather than DWAs. We made slight modifications to the rubric (which was used as a "prompt" to the model in this case) to enhance agreement with a set of human labels. Full agreement rates are given in Table \ref{tab:comparison}.
\end{itemize}

We construct three primary measures for our dependent variable of interest: (i) \textbf{$\alpha$}, corresponding to E1 in the exposure rubric above, anticipated to represent the lower bound of the proportion of exposed tasks within an occupation, (ii) \textbf{$\beta$}, which is the sum of E1 and 0.5*E2, where the 0.5 weight on E2 is intended to account for exposure when deploying the technology via complementary tools and applications necessitates additional investment, and (iii) \textbf{$\zeta$}, the sum of E1 and E2, an upper bound of exposure that provides an assessment of maximal exposure to an LLLM and LLM-powered software. We summarize agreement between annotation groups and measures in Table \ref{tab:comparison}. For the remainder of the analysis, if not specified, the reader may assume that we refer to $\beta$ exposure -- meaning all tasks directly exposed via tools like ChatGPT or the OpenAI Playground are considered twice as exposed as tasks requiring some complementary innovation.

\subsection{Limitations of our methodology}
\subsubsection{Subjective human judgments}
A fundamental limitation of our approach lies in the subjectivity of the labeling. In our study, we employ annotators who are familiar with LLM capabilities. However, this group is not occupationally diverse, potentially leading to biased judgments regarding LLMs' reliability and effectiveness in performing tasks within unfamiliar occupations. We acknowledge that obtaining high-quality labels for each task in an occupation requires workers engaged in those occupations or, at a minimum, possessing in-depth knowledge of the diverse tasks within those occupations. This represents an important area for future work in validating these results.

\subsubsection{Measuring LLMs with GPT-4}

\begin{figure}
    \centering
    \begin{minipage}{0.45\textwidth}
        \centering
        \includegraphics[width=\textwidth]{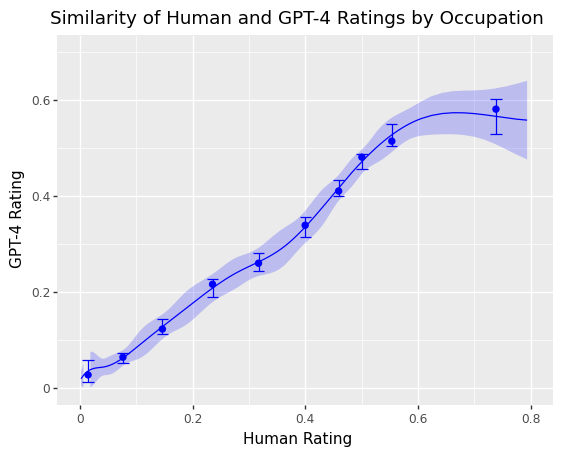}
    \end{minipage}\hfill
    \begin{minipage}{0.5\textwidth}
        \centering
        \includegraphics[width=\textwidth]{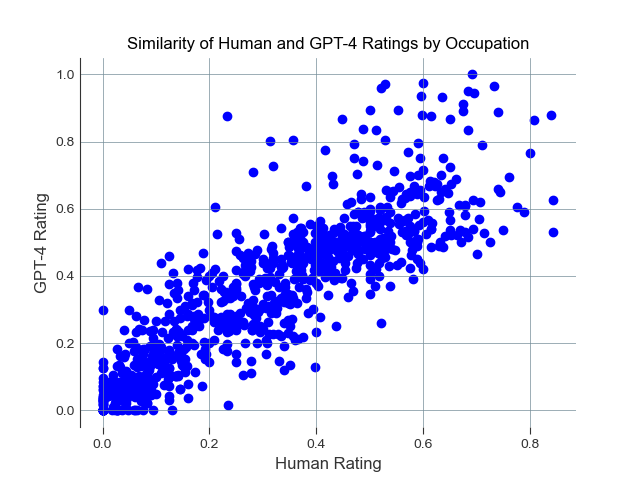}
    \end{minipage}
        \caption{Human raters (x-axis) and GPT-4 ratings (y-axis) show a high degree of agreement about LLM exposure by occupation. We compute occupation-level exposure in these figures by averaging the task-level exposures under the $\beta$ method. O*NET designates some tasks as "core" and others "supplemental". Core tasks are given twice the weight of supplemental tasks, and all weights sum to one. Near the highest levels of exposure following the $\beta$ method of aggregating exposure scores to occupations, GPT-4 ratings tend to be lower than Human ratings. We present the raw scatter plot and the binscatter. Near the top end of exposure ratings, humans are on average more likely to rate an occupation as exposed.}
        \label{fig:humangpt_exp_regscatter}
    \label{two_agreement_figs}    
\end{figure}

Recent research indicates that GPT-4 serves as an effective discriminator, capable of applying intricate taxonomies and responding to changes in wording and emphasis \citep{gpt4}. The outcomes of GPT-4 task classification are sensitive to alterations in the rubric's wording, the prompt's order and composition, the presence or absence of specific examples in the rubric, the level of detail provided, and the definitions given for key terms. Iterating on the prompt, based on observed outcomes in a small validation set, can enhance the agreement between model outputs and the rubric's intent. Consequently, there are slight differences between the rubric presented to humans and the one used for GPT-4. This decision was made deliberately to guide the model towards reasonable labels without excessively influencing human annotators. As a result, we use multiple annotation sources, but none should be considered the definitive ground truth relative to the others. In this analysis, we present results from human annotators as our primary results. Further improvement and innovation in crafting effective rubrics for LLM classification remains possible. Still, we observe a high degree of agreement between human ratings and GPT-4 ratings at the occupation level concerning overall exposure to LLM systems (see Table \ref{tab:comparison}, Figure \ref{two_agreement_figs}).

\subsubsection{Additional Weaknesses}

\begin{itemize}

\item \textbf{Validity of task-based framework.} It is unclear to what extent occupations can be entirely broken down into tasks, and whether this approach systematically omits certain categories of skills or tasks that are tacitly required for competent performance of a job. Additionally, tasks can be composed of sub-tasks, some of which are more automatable than others. Some tasks may function as pre-cursor to other tasks, such that the completion of downstream tasks is dependent on precursor tasks. If indeed, the task-based breakdown is not a valid representation of how most work in an occupation is performed, our exposure analysis would largely be invalidated.


\item \textbf{Lack of expertise and task interpretation.} Human annotators were mostly unaware of the specific occupations mapped to each DWA during the labeling process. This led to unclear logic for aggregating tasks and occupations, as well as some evident discrepancies in labels, demonstrated in Table \ref{tab:onet}. We experimented with various aggregation methods and discovered that even with a maximum-matching approach (taking the matching human<>model label if one existed), the agreement remained relatively consistent. Ultimately, we collected additional labels for task/occupation pairs where there was significant disagreement.

\item \textbf{Forward-looking and subject to change, with some early evidence.} Accurately predicting future LLM applications remains a significant challenge, even for experts \citep{gpt4}. The discovery of new emergent capabilities, changes in human perception biases, and shifts in technological development can all affect the accuracy and reliability of predictions regarding the potential impact of LLMs on worker tasks and the development of LLM-powered software. Our projections are inherently forward-looking and based on current trends, evidence, and perceptions of technological possibilities. As a result, they may change as new advancements arise in the field. For example, some tasks that seem unlikely for LLMs or LLM-powered software to impact today might change with the introduction of new model capabilities. Conversely, tasks that appear exposed might face unforeseen challenges limiting language model applications.

\item \textbf{Sources of disagreement.} While we did not rigorously examine sources of disagreement, we found a few places where humans and the model tended to get "stuck" in their assessments:
\begin{itemize}
    \item Tasks or activities where while an LLM could theoretically help or accomplish the task, adopting it to do so would require multiple people to change their habits or expectations (e.g. meetings, negotiations),
    \item Tasks or activities where there is currently some regulation or norm that requires or suggests human oversight, judgment or empathy (e.g. making decisions, counseling), and
    \item Tasks or activities where there already exists a technology that can reasonably automate the task (e.g. making reservations).
\end{itemize}
\end{itemize}

\section{Results}
\label{sec:econimpact}

General-purpose technologies are relatively rare and characterized by their pervasiveness, improvement over time, and the development of significant co-invention and spillovers \citep{lipsey2005economic}. Our assessment of LLMs' potential impact on the labor market is limited since it does not consider total factor productivity or capital input potential. In addition to their influence on labor, LLMs may also influence these dimensions.

At this stage, some general-purpose technology criteria are easier to evaluate than others. Our primary focus at this early stage is to test the hypothesis that LLMs have a pervasive influence on the economy, similar to the approach taken by \citep{goldfarb2023could}, who analyzed machine learning diffusion through job postings to assess its status as a general-purpose technology. Instead of using job postings or studying machine learning in general, we employ the task evaluation approach with both human and GPT-4 annotations. This analysis may reveal whether the impacts are limited to a specific set of similar tasks or occupations or if they will be more widespread.

Our findings suggest that, based on their task-level capabilities, LLMs have the potential to significantly affect a diverse range of occupations within the U.S. economy, demonstrating a key attribute of general-purpose technologies. In the following sections, we discuss results across various roles and wage structures. Additional results on the relative exposure of industries within the U.S. economy can be found in Appendix \ref{subsec:aggregates}.

\subsection{Summary Statistics}

Summary statistics for these measures can be found in Table \ref{tab:summarystats}. Both human and GPT-4 annotations indicate that average occupation-level $\alpha$ values fall between 0.14 and 0.15, suggesting that, on average, approximately 15\% of tasks within an occupation are directly exposed to LLMs.\footnote{We compute occupation-level scores for Table \ref{tab:summarystats} assigning double the weight to tasks designated as "core" by O*NET as tasks designated "supplemental". All tasks weights sum to one within an occupation.} This figure increases to over 30\% for $\beta$ and surpasses 50\% for $\zeta$. Coincidentally, human and GPT-4 annotations also tag between 15\% and 14\% of total tasks in the dataset as being exposed to LLMs. Based on the $\beta$ values, we estimate that 80\% of workers belong to an occupation with at least 10\% of its tasks exposed to LLMs, while 19\% of workers are in an occupation where over half of its tasks are labeled as exposed.

We ran one set of analyses using O*NET's "Importance" scores but did not find significant changes to our findings. Though we do acknowledge that not weighting relative importance of a task to a given occupation yields some curious results (e.g. ranking Barbers as having reasonably high exposure).

Although the potential for tasks to be affected is vast, LLMs and LLM-powered software must be incorporated into broader systems to fully realize this potential. As is common with general-purpose technologies, co-invention barriers may initially impede the rapid diffusion of GPTs into economic applications. Furthermore, predicting the need for human oversight is challenging, especially for tasks where model capabilities equal or surpass human levels. While the requirement for human supervision may initially slow down the speed at which these systems diffuse through the economy, users of LLMs and LLM-powered systems are likely to become increasingly acquainted with the technology over time, particularly in terms of understanding when and how to trust its outputs.

\input{tables/summary_stats}
\subsection{Wages and Employment}
\label{subsec:labor}

\begin{figure}
    \centering
    
     \includegraphics[width=.5\textwidth]{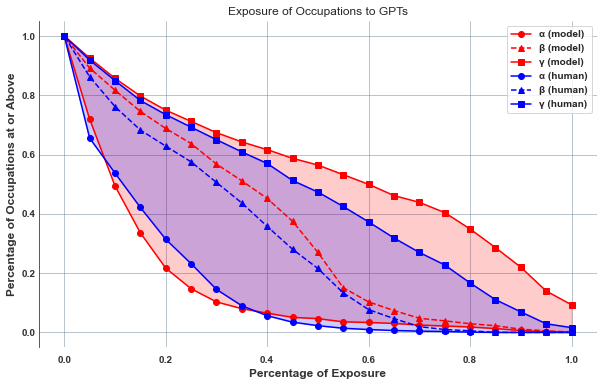}
    \caption{Exposure intensity across the economy, displayed in terms of percent of affected occupations. A given data point gives the percent of occupations with exposure below the given threshold. A previous version of this paper had two labels reversed in the chart, flipping human and model responses. In this figure, all tasks within an occupation are given equal weight.}
\label{fig:exposure_final}
\end{figure}

In Figure \ref{fig:exposure_final}, we present exposure intensity across the economy. Each point on the plot displays occupational exposure, where the point's x-axis value represents the share of an occupation's tasks that are exposed (at each level $\alpha$, $\beta$, and $\zeta$) and the point's y-axis value represents the share of all US occupations with that share of tasks exposed. For example, human annotators determined that 2.3\% of occupations are $\alpha_{50}$-exposed, 21.6\% are $\beta_{50}$-exposed, and 47.3\% are $\zeta_{50}$-exposed, where the threshold of 50\% comes from the x-axis and the percentage of occupations comes from the y axis. At any given point on the x-axis, the vertical distance between the $\alpha$ and the $\zeta$ represents the exposure potential attributable to tools and applications beyond direct exposure to LLMs. All tasks within an occupation in this figure are given equal weight. 

\begin{figure}[htbp]\small
\centering
\begin{minipage}[t]{0.48\textwidth}
\centering
\includegraphics[width=\linewidth]{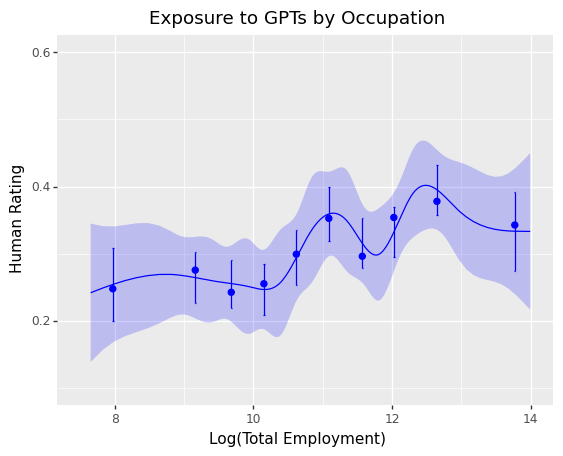}
\label{fig:meanhuman_totemp}
\end{minipage}
\hfill
\begin{minipage}[t]{0.48\textwidth}
\centering
\includegraphics[width=\linewidth]{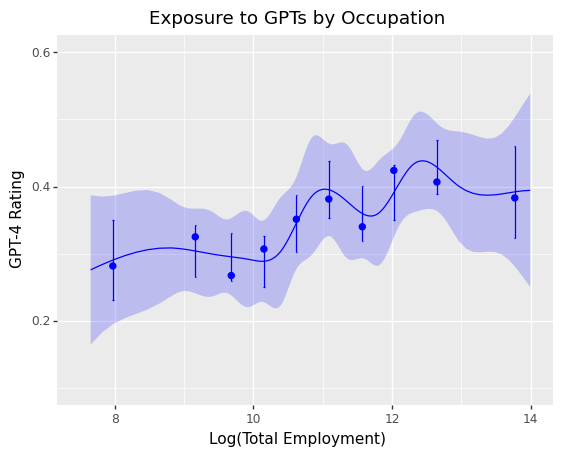}
\label{fig:dve_totemp}
\end{minipage}
\begin{minipage}[t]{0.48\textwidth}
\centering
\includegraphics[width=\linewidth]{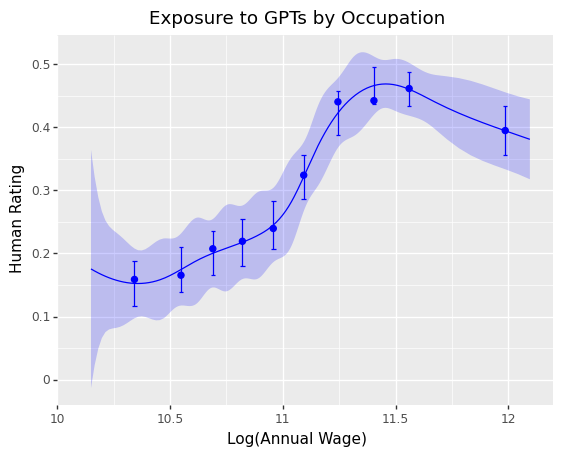}
\label{fig:human_binscatter}
\end{minipage}
\hfill
\begin{minipage}[t]{0.48\textwidth}
\centering
\includegraphics[width=\linewidth]{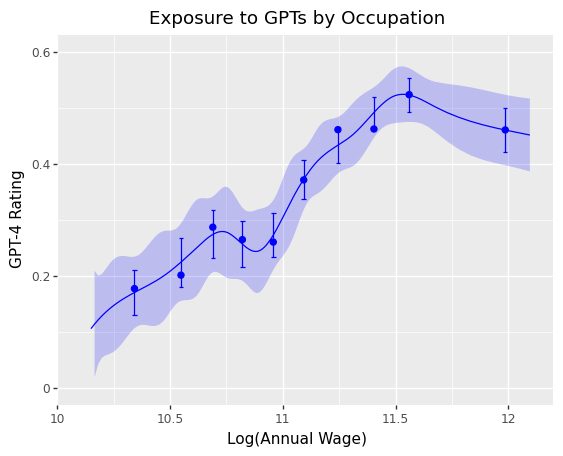}
\label{fig:dve_binscatter}
\end{minipage}
  \caption{The binscatter plots depict the exposure to language models (LLMs) in various occupations, as assessed by both human evaluators and GPT-4. These plots compare the exposure to LLM and partial LLM-powered software ($\beta$) at the occupation level against the log of total employment within an occupation and log of the median annual wage for occupations. While some discrepancies exist, both human and GPT-4 assessments indicate that higher wage occupations tend to be more exposed to LLMs. Additionally, numerous lower wage occupations demonstrate high exposure based on our rubric. Core tasks receive twice the weight of supplemental tasks within occupations when calculating average exposure scores. Employment and wage data are sourced from the BLS-OES survey conducted in May 2021. In aggregating tasks to the occupation-level, we assign half the weight to O*NET supplemental tasks as we do for core tasks. All weights within an occupation sum to one.}
  \label{binscatters}
\end{figure}

Aggregated at the occupation level, human and GPT-4 annotations exhibit qualitative similarities and tend to correlate, as demonstrated in Figure \ref{binscatters}. Human annotations estimate marginally lower exposure for high-wage occupations compared to GPT-4 annotations. While there are numerous low-wage occupations with high exposure and high-wage occupations with low exposure, the overall trend in the binscatter plot reveals that higher wages are associated with increased exposure to LLMs.\footnote{In aggregating tasks to the occupation-level, we assign half the weight to O*NET supplemental tasks as we do for core tasks.}

\input{tables/top_jobs_sample}

The potential exposure to LLMs seems to have little correlation with current employment levels. In Figure \ref{binscatters}, both human and GPT-4 ratings of overall exposure are aggregated to the occupation-level (y-axis) and compared with the log of total employment (x-axis). Neither plot reveals significant differences in LLM exposure across varying employment levels.

\subsection{Skill Importance} 

In this section, we explore the relationship between the importance of a skill for an occupation (as annotated in the O*NET dataset) and our exposure measures. First, we use the Basic Skills provided by O*NET (skill definitions can be found in Appendix \ref{sec:skills_definitions}) and normalize the measure of skill importance for each occupation to improve the comprehensibility of the results. Next, we conduct a regression analysis on our exposure measures ($\alpha$, $\beta$, $\zeta$) to examine the strength of associations between skill importance and exposure.

Our findings indicate that the importance of \textbf{science} and \textbf{critical thinking} skills are strongly negatively associated with exposure, suggesting that occupations requiring these skills are less likely to be impacted by current LLMs. Conversely, \textbf{programming} and \textbf{writing} skills show a strong positive association with exposure, implying that occupations involving these skills are more susceptible to being influenced by LLMs (see Table \ref{tab:skills} for detailed results). 

\input{tables/skill_correlation}

\subsection{Barriers to Entry}

Next, we examine barriers to entry to better understand if there is differentiation in exposure due to types of jobs. One such proxy is an O*NET occupation-level descriptor called the "Job Zone." \href{https://www.onetonline.org/help/online/zones}{A Job Zone} groups occupations that are similar in (a) the level of education needed to get a job in the occupation, (b) the amount of related experience required to do the work, and (c) the extent of on-the-job training needed to do the work. In the O*NET database, there are 5 Job Zones, with Job Zone 1 requiring the least amount of preparation (3 months) and Job Zone 5 requiring the most extensive amount of preparation, 4 or more years. We observe that median income increases monotonically across Job Zones as the level of preparation needed also increases, with the median worker in Job Zone 1 earning $\$30,230$ and the median worker in Job Zone 5 earning $\$80,980$.

\begin{figure}
    \centering
    \includegraphics[width=.95\textwidth]{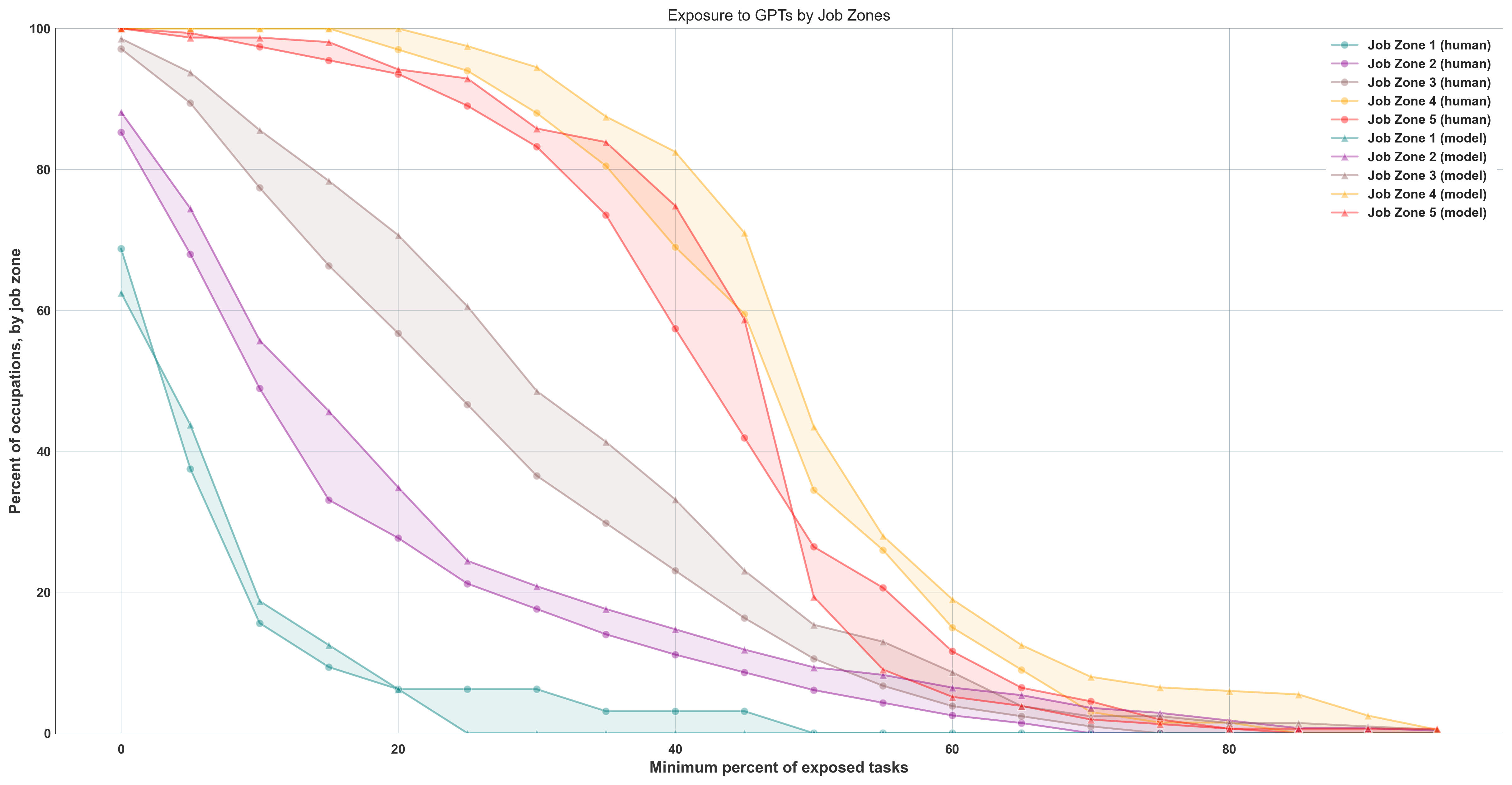}
    \caption{$\beta$ exposure ratings of occupations in the five Job Zones, which are groups of similar occupations that are classified according to the level of education, experience, and on-the-job training needed to perform them. All tasks are weighted equally.}
    \label{fig:job_zone_plots}
\end{figure}

All of our measures ($\alpha$, $\beta$, and $\zeta$) show an identical pattern, that is, exposure increases from Job Zone 1 to Job Zone 4, and either remains similar or decreases at Job Zone 5. Similar to Figure \ref{fig:exposure_final}, in Figure \ref{fig:job_zone_plots}, we plot the percentage of workers at every threshold of exposure. We find that, on average, the percentage of workers in occupations with greater than 50\% $\beta$ exposure in Job Zones 1 through 5 have $\beta$ at 0.00\% (Job Zone 1), 6.11\% (Job Zone 2), 10.57\% (Job Zone 3), 34.5\% (Job Zone 4), and 26.45\% (Job Zone 5), respectively.\footnote{For this set of results, all tasks have equal weight within an occupation. Results do not change meaningfully with the core/supplemental weighting scheme.} 

\input{tables/job_zone_summary}

\label{table:job_zone_table}

\subsubsection{Typical Education Needed for Entry}

Since inclusion in a Job Zone accounts for both the education required—which itself is a proxy for skill acquisition—and the preparation required, we seek data to disentangle these variables. We use two variables from the Bureau of Labor Statistics' Occupational data: "Typical Education Needed for Entry" and "On-the-job Training Required to Attain Competency" in an occupation. By examining these factors, we aim to uncover trends with potential implications for the workforce. There are 3,504,000 workers for whom we lack data on education and on-the-job training requirements, and they are therefore excluded from the summary tables.

Our analysis suggests that individuals holding Bachelor's, Master's, and professional degrees are more exposed to LLMs and LLM-powered software than those without formal educational credentials (see Table \ref{tab:on_the_job_training}). Interestingly, we also find that individuals with some college education but no degree exhibit a high level of exposure to LLMs and LLM-powered software. Upon examining the table displaying barriers to entry, we observe that the jobs with the least exposure require the most training, potentially offering a lower payoff (in terms of median income) once competency is achieved. Conversely, jobs with no on-the-job training required or only internship/residency required appear to yield higher income but are more exposed to LLMs.

\include{tables/on_the_job_training}

\section{Validation of Measures}
\label{sec:validation}

\subsection{Comparison to Earlier Efforts}
\label{subsec:othermeasures}
This paper aims to build on a number of previous empirical studies examining the occupational exposure to advances in AI and/or automation. Previous studies have used a variety of methods, including:
\begin{itemize}
    \item Using occupational taxonomies like O*NET to characterize which occupations have routine vs. non-routine and manual vs. cognitive task content \citep{autor2003skill, acemoglu2011skills}.
    \item Mapping text descriptions of tasks to descriptions of technological advances in patents. \citep{NBERw29552, Webb2020}
    \item Linking capabilities of AI systems to occupational abilities and aggregating exposure estimates to the occupations where those abilities are required. \citep{SeamansRajFelten2018, felten2023will}
    \item Mapping the results of AI task benchmark evaluations (ImageNet, Robocup, etc.) to 59 worker tasks through a set of 14 cognitive abilities drawn from the cognitive science literature. \citep{Tolan2021}
    \item Expert labeling of automation potential for a set of O*NET occupations where experts had high confidence, combined with a probabilistic classifier to estimate automation potential for the remainder of O*NET occupations. \citep{FreyOsborne2017}
    \item Developing a rubric for evaluating the "suitability for machine learning" (SML) of activities that workers are completing in the economy \citep{brynjolfsson2017can,Brynjolfsson2018, brynjolfssonQuantifyingDistributionMachine2023}. 
\end{itemize}

We provide a set of summary statistics on many of these prior efforts in Table \ref{tab:autoscores_sumstats}. 

This paper's methodology primarily builds upon the SML approach by developing a rubric to evaluate the overlap between LLM capabilities and worker tasks as reported in the O*NET database. Table \ref{tab:autoscores_regression} presents the results of OLS regressions of our new LLM exposure measurements on occupation-level exposure measures from \citep{SeamansRajFelten2018} ("AI Occupational Exposure Score" in the table), \citep{FreyOsborne2017} (Frey \& Osborne Automation), scores from all three technologies in \citep{Webb2020}, normalized routine manual and cognitive scores from \citep{acemoglu2011skills}, and \citep{Brynjolfsson2018,brynjolfssonQuantifyingDistributionMachine2023} (SML). We also use annualized occupational salaries from the most recent BLS Occupational Employment Survey as a control. There are four separate output variables representing new scores in this paper that are predicted by earlier efforts.

GPT-4 Exposure Rating 1 corresponds to our overall exposure rubric as evaluated by GPT-4, where full exposure potential is coded as 1, no exposure potential is coded as 0, and partial exposure (E2 in our labeling scheme) is coded as 0.5. GPT-4 Exposure Rating 2 is scored similarly for overall exposure, but with a slightly different prompt. The results are very similar across the two prompts. Human Exposure Rating represents the same rubric as in GPT-4 Exposure Rating 1 but is scored by humans, as discussed in an earlier section of the paper. These results correspond to the $\beta$ set of statistics presented above, with supplemental tasks having half the weight of core tasks within an occupation. These weights sum to one (core/supplemental distinctions are determined by O*NET).

The results across each type of measurement are consistent. We find generally positive and statistically significant correlations between our LLM exposure measures and previous measurements targeting software and AI. Encouragingly, the SML exposure scores by occupation show significant and positive associations with the exposure scores we develop in this paper, demonstrating a level of cohesion between the two studies with similar approaches. The Webb software and AI patent-based measures, SML, and normalized (demeaned and divided by standard deviation) routine cognitive scores all exhibit positive associations with some of our measures.

Software, SML, and routine cognitive scores all show positive and statistically significant associations with LLM exposure scores at a 1\% level. Coefficients on AI scores from \citep{Webb2020} are also positive and statistically significant at a 5\% level, but our secondary prompt on overall exposure to LLMs in columns 3 and 4 does not exhibit a statistically significant relationship. For the most part, the AI Occupational Exposure Score is not correlated with our exposure measures. Webb's Robot exposure scores, routine manual task content, and the overall Automation metric from \citep{FreyOsborne2017} are all negatively correlated with our primary GPT-4 and human-assessed overall exposure ratings, conditional on the other measurements. This negative correlation reflects the limited exposure of physical tasks to LLMs. Manual work is not exposed to LLMs or even LLMs with additional systems integration for the time being. 

Low correlations with \citep{SeamansRajFelten2018} and \citep{FreyOsborne2017} could potentially be explained by differences in approaches. Linking AI capabilities to worker abilities or scoring exposure directly based on the occupation's characteristics, rather than aggregating up to the occupation from DWA or task-level scoring (as in the SML paper and our own), offer a slightly different perspective on the content of occupations.

In all regressions, the $R^2$ ranges between 60.7\% (column 3) and 72.8\% (column 5). This suggests that our measure, which explicitly focuses on LLM capabilities, has between 28 and 40\% unexplained variance compared to other measurements. Particularly in the case of AI-related exposure scores, we anticipate that a combination of other measurements would have a strong correlation with our scores. However, earlier efforts had limited information about the future progress of LLMs or LLM-powered software. We expect that our understanding of future machine learning technologies is similarly imperfectly captured by our rubric today.

\begin{table}[]\small
\begin{threeparttable}
    \centering
    \resizebox{\textwidth}{!}{
    \input{tables/autoscores_summary}
    }
    \caption{Summary statistics for a suite of prior efforts to measure occupational exposure to AI and automation. We have also included summary statistics for measurements newly presented in this work. We include all measures from \citep{Webb2020}, normalized routine cognitive and manual scores from \citep{acemoglu2011skills} (means may deviate slightly from 0 due to imperfect matching of occupational groups), Suitability for Machine Learning from \citep{brynjolfsson2017can, Brynjolfsson2018, brynjolfssonQuantifyingDistributionMachine2023}, AI Occupational Exposure from \citep{SeamansRajFelten2018}, and Automation exposure from \citep{FreyOsborne2017}. We include as many occupations as we can match, but since O*NET taxonomies have changed as these measures have been developed, some of the roles may be missing from the most recent version of O*NET 6-digit occupations.}
    \label{tab:autoscores_sumstats}
\end{threeparttable}
\end{table}

\begin{table}[]\small
\begin{threeparttable}
    \centering
    \resizebox{\textwidth}{!}{
    \input{tables/autoscores_regression}
    }
    \caption{Regression of LLM-exposure scores on prior measures of occupational exposure to AI and automation. We also include annualized wages from the BLS-OES survey in May 2021. Each measure is kept in its original scale, with the exception of routine cognitive and routine manual scores from \citep{acemoglu2011skills}. Those two scores are standardized to mean zero and variance 1. Generally we find strong positive associations with previous efforts, though large residual variance to still be explained by our new measures. Columns 1 and 2 are based on our main $\beta$ exposure measure from GPT-4 ratings. Columns 3 and 4 are based on a similar slightly different exposure rubric also rated by GPT-4 for robustness. Columns 5 and 6 reflect human ratings on the same rubric as columns 1 and 2. Occupation-level scores are built using the core/supplemental task weights, assigning supplemental tasks as having half the weight of core tasks.}
    \label{tab:autoscores_regression}
\end{threeparttable}
\end{table}

\section{Discussion}
\label{sec:discussion}

\subsection{GPTs as a General-Purpose Technology}

Earlier in this paper we discuss the possibility that LLMs could be classified as a general-purpose technology. This classification requires LLMs to meet three core criteria: improvement over time, pervasiveness throughout the economy, and the ability to spawn complementary innovations \citep{lipsey2005economic}. Evidence from the AI and machine learning literature thoroughly demonstrates that LLMs meet the first criteria -- they are improving in capabilities over time with the ability to complete or be helpful for an increasingly complex set of tasks and use-cases (see \ref{sec:llmlit}). This paper presents evidence to support the latter two criteria, finding that LLMs on their own can have pervasive impacts across the economy, and that complementary innovations enabled by LLMs -- particularly via software and digital tools -- can have widespread application to economic activity.

Figure \ref{fig:exposure_final} offers one illustration of the potential economic impact of complementary software built on top of LLMs. Taking the difference in the y-axis (the share of all occupations) between $\alpha$ and $\zeta$ at a given point along the x-axis (the share of tasks within an occupation that are exposed) gives the aggregate within-occupation exposure potential attributable to tools and software over and above direct exposure from LLMs on their own. The difference in means across all tasks between $\alpha$ and $\zeta$ of 0.42 using the GPT-4 annotations and 0.32 using the human annotations (see Figure \ref{tab:summarystats}), suggests that the average impact of LLM-powered software on task-exposure may be more than twice as large as the mean exposure from LLMs on their own (mean $\zeta$ of 0.14 based on both human annotations and GPT-4 annotations). While our findings suggest that out-of-the-box these models are relevant to a meaningful share of workers and tasks, they also suggest that the software innovations they spawn could drive a much broader impact.

One component of the pervasiveness of a technology is its level of adoption by businesses and users. This paper does not systematically analyze adoption of these models, however, there is early qualitative evidence that adoption and use of LLMs is becoming increasingly widespread. The power of relatively simple UI improvements on top of LLMs was evident in the rollout of ChatGPT -- wherein versions of the underlying language model had been previously available via API, but usage skyrocketed after the release of the ChatGPT interface. \citep{chow_chatgpt_2023, chatgptblog} Following this release, a number of commercial surveys indicate that firm and worker adoption of LLMs has increased over the past several months. \citep{constantz_bloomberg, resumebuildersuvey}

Widespread adoption of these models requires addressing existing bottlenecks. A key determinant of their utility is the level of confidence humans place in them and how humans adapt their habits. For instance, in the legal profession, the models' usefulness depends on whether legal professionals can trust model outputs without verifying original documents or conducting independent research. The cost and flexibility of the technology, worker and firm preferences, and incentives also significantly influence the adoption of tools built on top of LLMs. In this way, adoption may be driven by progress on some of the ethical and safety risks associated with LLMs: bias, fabrication of facts, and misalignment, to name a few \cite{4systemcard}. Moreover, the adoption of LLMs will vary across different economic sectors due to factors such as data availability, regulatory environment, and the distribution of power and interests. Consequently, a comprehensive understanding of the adoption and use of LLMs by workers and firms requires a more in-depth exploration of these intricacies.

One possibility is that time savings and seamless application will hold greater importance than quality improvement for the majority of tasks. Another is that the initial focus will be on augmentation, followed by automation \citep{huang2018artificial}. One way this might take shape is through an augmentation phase where jobs first become more precarious (e.g., writers becoming freelancers) before transitioning to full automation.

\subsection{Implications for US Public Policy}

The introduction of automation technologies, including LLMs, has previously been linked to heightened economic disparity and labor disruption, which may give rise to adverse downstream effects \citep{acemoglu2022demographics, Acemoglu2002, Moll2021, Klinova2021, Weidinger2021, Weidinger2022}. Our results examining worker exposure in the United States underscore the need for societal and policy preparedness to the potential economic disruption posed by LLMs and the complementary technologies that they spawn. While it is outside the scope of this paper to recommend specific policy prescriptions to smooth the transition to an economy with increasingly widespread LLM adoption, prior work such as \citep{Autor2022} has articulated several important directions for US policy related to education, worker training, reforms to safety net programs, and more. 

\subsection{Limitations and Future Work}

In addition to those discussed above, we highlight some particular limitations of this work that warrant further investigation. Primarily, our focus on the United States restricts the generalizability of our findings to other nations where the adoption and impact of generative models may differ due to factors such as industrial organization, technological infrastructure, regulatory frameworks, linguistic diversity, and cultural contexts. We hope to address this limitation by extending the study's scope and by sharing our methods so other researchers can build on them.

Subsequent research efforts should consider two additional studies: one exploring LLM adoption patterns across various sectors and occupations, and another scrutinizing the actual capabilities and limitations of state-of-the-art models in relation to worker activities beyond the scope of our exposure scores. For example, despite recent advances in multimodal capabilities with GPT-4, we did not consider vision capabilities in the $\alpha$ ratings on direct LLMs-exposure \citep{gpt4}. Future work should consider the impact of such capability advances as they unfold. Furthermore, we acknowledge that there may be discrepancies between theoretical and practical performance, particularly in complex, open-ended, and domain-specific tasks.

\section{Conclusion}
\label{sec:conclusion}

In conclusion, this study offers an examination of the potential impact of LLMs on various occupations and industries within the U.S. economy. By applying a new rubric for understanding LLM capabilities and their potential effects on jobs, we have observed that most occupations exhibit some degree of exposure to LLMs, with higher-wage occupations generally presenting more tasks with high exposure. Our analysis indicates that approximately 19\% of jobs have at least 50\% of their tasks exposed to LLMs when considering both current model capabilities and anticipated LLM-powered software. 

Our research aims to highlight the general-purpose potential of LLMs and their possible implications for US workers. Previous literature demonstrates the impressive improvements of LLMs to date (see \ref{sec:llmlit}). Our findings confirm the hypothesis that these technologies can have pervasive impacts across a wide swath of occupations in the US, and that additional advancements supported by LLMs, mainly through software and digital tools, can have significant effects on a range of economic activities. However, while the technical capacity for LLMs to make human labor more efficient appears evident, it is important to recognize that social, economic, regulatory, and other factors will influence actual labor productivity outcomes. As capabilities continue to evolve, the impact of LLMs on the economy will likely persist and increase, posing challenges for policymakers in predicting and regulating their trajectory.

Further research is necessary to explore the broader implications of LLM advancements, including their potential to augment or displace human labor, their impact on job quality, impacts on inequality, skill development, and numerous other outcomes. By seeking to understand the capabilities and potential effects of LLMs on the workforce, policymakers and stakeholders can make more informed decisions to navigate the complex landscape of AI and its role in shaping the future of work.

\subsection{LLM Conclusion (GPT-4's Version)}

Generative Pre-trained Transformers (GPTs) generate profound transformations, garnering potential technological growth, permeating tasks, greatly impacting professions. This study probes GPTs’ potential trajectories, presenting a groundbreaking rubric to gauge tasks’ GPT exposure, particularly in the U.S. labor market.

\subsection{LLM Conclusion (Author-Augmented Version)}

Generative Pre-trained Transformers (GPTs) generate profound transformations, garnering potential technological growth, permeating tasks, gutting professional management. Gauging possible trajectories? Generate pioneering taxonomies, gather policymakers together, generalize past today.

\section*{Acknowledgments}
Thank you to the group of annotators who helped us annotate task exposure, including Muhammad Ahmed Saeed, Bongane Zitha, Merve Özen Şenen, J.J., and Peter Hoeschele. We also thank Lauryn Fuld, Ashley Glat, Michael Lampe, and Julia Susser for excellent research assistance. We thank Miles Brundage for significant feedback on this paper.

We thank Todor Markov and Vik Goel for setting up the infrastructure we use to run our rubrics with GPT-4. We thank Lama Ahmad, Donald Bakong, Seth Benzell, Erik Brynjolfsson, Parfait Eloundou-Enyegue, Carl Frey, Sarah Giroux, Gillian Hadfield, Johannes Heidecke, Alan Hickey, Eric Horvitz, Shengli Hu, Ashyana Kachra, Christina Kim, Katya Klinova, Daniel Kokotajlo, Gretchen Krueger, Michael Lampe, Aalok Mehta, Larissa Schiavo, Daniel Selsam, Sarah Shoker, Prasanna Tambe, and Jeff Wu for feedback and edits at various stages of the project.

\section*{LLM assistance statement}
GPT-4 and ChatGPT were used for writing, coding, and formatting assistance in this project.

\appendix

\section{Rubric}\label{taxonomies}

\subsection{Exposure} \label{exposure_tax} 
\input{exposure_taxonomy}

\section{O*NET Basic Skills Definitions}
\label{sec:skills_definitions}

\subsection*{Basic Skills}

Developed capacities that facilitate learning or the more rapid acquisition of knowledge.

\subsection*{Content}

Background structures needed to work with and acquire more specific skills in a variety of different domains.

\begin{itemize}
    \item \textbf{Reading Comprehension} — Understanding written sentences and paragraphs in work-related documents.
    \item \textbf{Active Listening} — Giving full attention to what other people are saying, taking time to understand the points being made, asking questions as appropriate, and not interrupting at inappropriate times.
    \item \textbf{Writing} — Communicating effectively in writing as appropriate for the needs of the audience.
    \item \textbf{Speaking} — Talking to others to convey information effectively.
    \item \textbf{Mathematics} — Using mathematics to solve problems.
    \item \textbf{Science} — Using scientific rules and methods to solve problems.
\end{itemize}

\subsection*{Process}

Procedures that contribute to the more rapid acquisition of knowledge and skill across a variety of domains

\begin{itemize}
    \item \textbf{Critical Thinking} — Using logic and reasoning to identify the strengths and weaknesses of alternative solutions, conclusions or approaches to problems.
    \item \textbf{Active Learning} — Understanding the implications of new information for both current and future problem-solving and decision-making.
    \item \textbf{Learning Strategies} — Selecting and using training/instructional methods and procedures appropriate for the situation when learning or teaching new things.
    \item \textbf{Monitoring} — Monitoring/Assessing performance of yourself, other individuals, or organizations to make improvements or take corrective action.
\end{itemize}

\subsection*{Cross-Functional Skills}
Note: We selected only Programming from the list of cross-functional skills because of our prior knowledge about the models' ability to code.
\begin{itemize}
    \item \textbf{Programming} - Writing computer programs for various purposes.
\end{itemize}

\begin{table}[h]
    \centering
    \input{tables/education_table}
    \caption{Mean exposure scores for occupations, grouped by typical education needed for entry into the occupation. Alongside exposure scores, we display the median of median annual income for each occupation, as well as the total number of workers in each group, in thousands.}
    \label{tab:education}
\end{table}

\section{Industrial and Productivity Exposure}
\label{subsec:aggregates}

\begin{figure}[p]
\centering
\includegraphics[width=\textheight, angle=270]{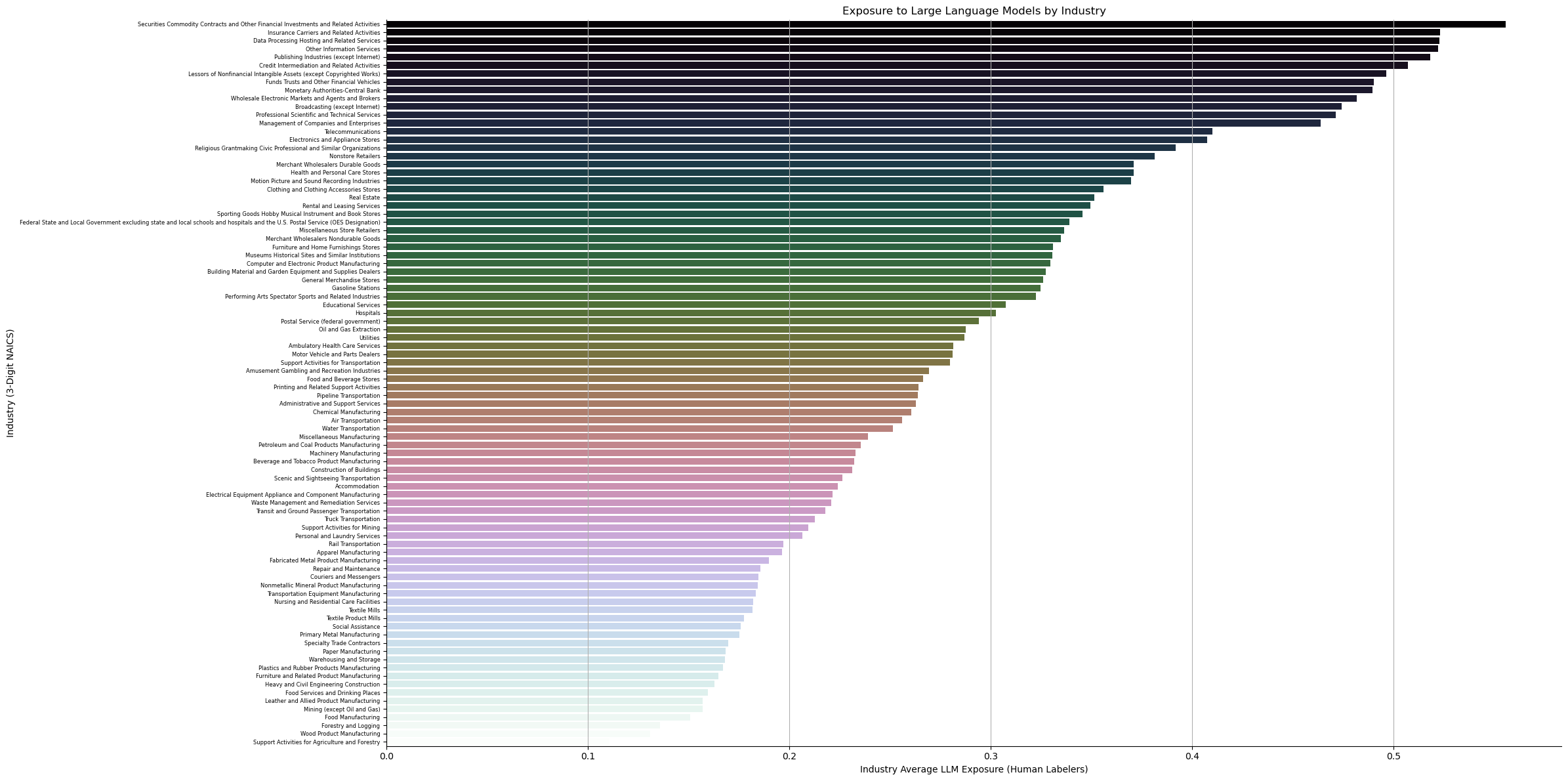}
\caption{}
\label{fig:meanhuman_indexp}
\end{figure}

\begin{figure}[p]
\centering
\includegraphics[width=\textheight, angle=270]{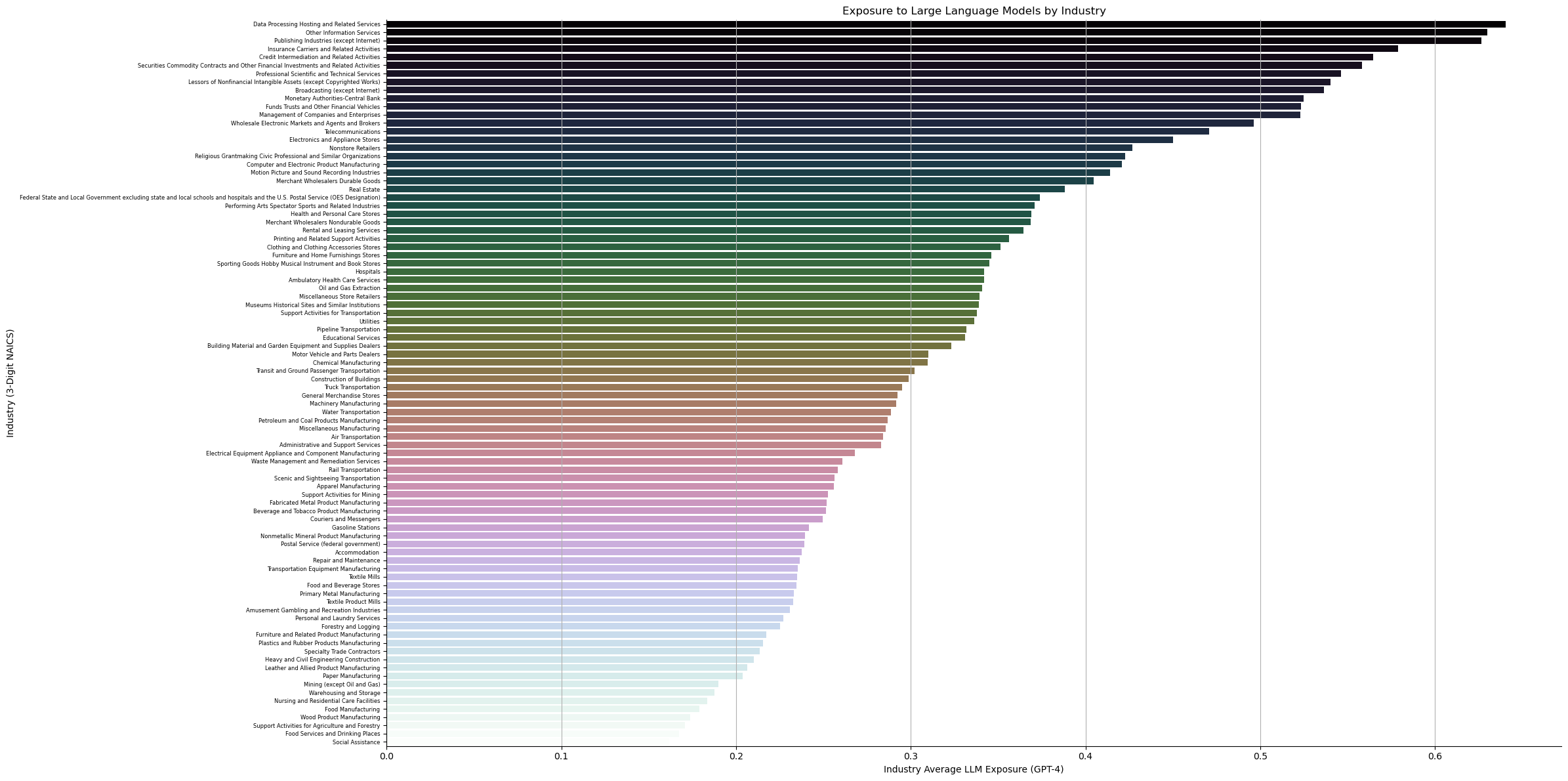}
\caption{}
\label{fig:dve_indexp}
\end{figure}

Figures \ref{fig:meanhuman_indexp} and \ref{fig:dve_indexp} show the overall employment-weighted relative exposure of 3-digit NAICS industries according to human raters and GPT-4 respectively (based on our exposure rubric). The impact potential is present across nearly all industries, with wide heterogeneity. Both methods agree generally on relative exposures: data processing, information processing, and hospitals all have high exposure.\footnote{Aggregations are done according to the $\beta$ method described above. Task weights for supplemental tasks are half the core task weights, summing to one within an occupation. Occupation counts are used as weights to build industry-level measures of exposure.}

\includegraphics[width=0.5\textwidth]{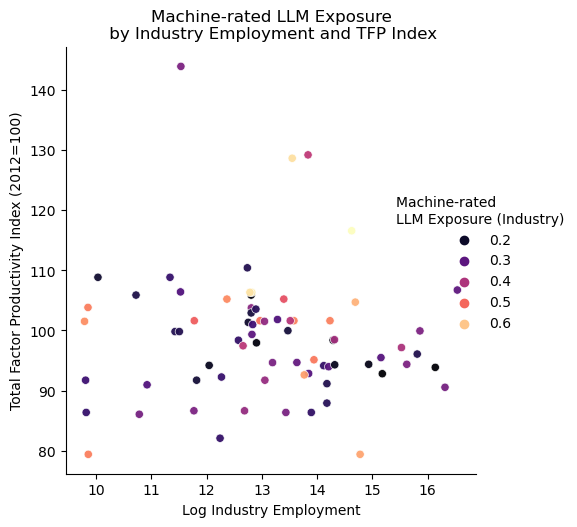}
\label{fig:tfp_dv}
\includegraphics[width=0.5\textwidth]{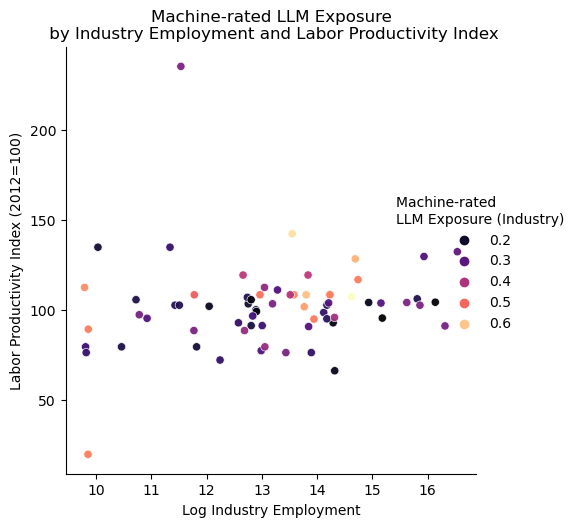}
\label{fig:lp_dv}
Recent productivity growth (both total factor and labor) appears uncorrelated with exposure as well. Figures \ref{fig:tfp_dv} and \ref{fig:lp_dv} show little relationship between productivity growth since 2012 and current exposure to LLMs as rated by the model. A high correlation between already fast-growing productive industries and exposure might mean an exacerbation of Baumol's cost disease. In other words, if LLMs are likely to increase productivity differentially across industries, one concern is that the most productive would become even more productive. With inelastic demand for the production of those industries, the most productive sectors would shrink as a proportion of inputs in the economy. We see little to suggest this will be the case. Productivity growth since 2012 and exposure to LLM technologies appear unrelated.\footnote{As above, aggregations are done according to the $\beta$ method described above. Task weights for supplemental tasks are half the core task weights, summing to one within an occupation. Occupation counts are used as weights to build industry-level measures of exposure.}

\section{Occupations Without Any Exposed Tasks}
\label{subsec:noexposure}
\input{tables/bottom_jobs}


\bibliographystyle{apalike} 

\bibliography{main}

\end{document}

%% file: tables/onet_tables.tex
\begin{table}\scriptsize \centering
\begin{tabular}{>{\raggedleft\arraybackslash}p{1cm}>{\raggedright\arraybackslash}p{2.4cm}>{\raggedright\arraybackslash}p{4.5cm}>{\raggedright\arraybackslash}p{5.2cm}}
\toprule
\rowcolor{gray!20}
{\raggedright\arraybackslash}\textbf{Task ID}  &  \textbf{Occupation Title}  &  \textbf{DWAs}  &  \textbf{Task Description} \\
\midrule 
14675 & Computer Systems Engineers/Architects & Monitor computer system performance to ensure proper operation. & Monitor system operation to detect potential problems.\\
\rowcolor{gray!10}
18310  &  Acute Care Nurses  &  Operate diagnostic or therapeutic medical instruments or equipment.\newline Prepare medical supplies or equipment for use.   &  Set up, operate, or monitor invasive equipment and devices, such as colostomy or tracheotomy equipment, mechanical ventilators, catheters, gastrointestinal tubes, and central lines.\\ 
4668.0 & Gambling Cage Workers & Execute sales or other financial transactions. & Cash checks and process credit card advances for patrons.\\
\rowcolor{gray!10}
15709  &  Online Merchants  &  Execute sales or other financial transactions.  &  Deliver e-mail confirmation of completed transactions and shipment. \\ 
6529 & Kindergarten Teachers, Except Special Education & -- & Involve parent volunteers and older students in children's activities to facilitate involvement in focused, complex play. \\
\rowcolor{gray!10}

6568 & Elementary School Teachers, Except Special Education & --  & Involve parent volunteers and older students in children's activities to facilitate involvement in focused, complex play. \\  \bottomrule
\end{tabular}
\caption{Sample of occupations, tasks, and Detailed Work Activities from the O*NET database. We see that aggregating over activities alone is imprecise, as evidenced by the fact that we'd expect Gambling Cage Workers to complete the given DWA in person, using some physicality while we'd expect Online Merchants to complete the same activity solely with a computer.}
\label{tab:onet}
\end{table}

%% file: tables/summary_stats.tex
\begin{table}[h!]\small\centering
\begin{tabular}{l c c c c}
  \toprule
  \multicolumn{5}{c}{\bfseries Occupation Level Exposure} \\
  \cmidrule{1-5}
  & \multicolumn{2}{c}{\bfseries Human} & \multicolumn{2}{c}{\bfseries GPT-4} \\
  \cmidrule(lr){2-3} \cmidrule(lr){4-5}
   & mean & std & mean & std \\
$\pmb{\alpha}$ & 0.14 & 0.14 & 0.14 & 0.16  \\
$\pmb{\beta}$ & 0.30 & 0.21 & 0.34 & 0.22 \\
$\pmb{\zeta}$ & 0.46 & 0.30 & 0.55 & 0.34 \\
  \midrule
  \multicolumn{5}{c}{\bfseries Task Level Exposure} \\
  \cmidrule{1-5}
  & \multicolumn{2}{c}{\textbf{Human}} & \multicolumn{2}{c}{\textbf{GPT-4}} \\
  \cmidrule(lr){2-3} \cmidrule(lr){4-5}
  & mean & std & mean & std \\
  $\pmb{\alpha}$ & 0.15 & 0.36 & 0.14 & 0.35 \\
  $\pmb{\beta}$ & 0.31 & 0.37 & 0.35 & 0.35 \\
  $\pmb{\zeta}$ & 0.47  & 0.50 & 0.56 & 0.50 \\
  \bottomrule
\end{tabular}
\caption{Summary statistics of our human and model exposure data. Tasks designated as core tasks for an occupation are given twice the weight as those indicated to be supplemental in the O*NET task file.}
\label{tab:summarystats}
\end{table}
  

%% file: tables/top_jobs_sample.tex
\begin{table}[h]\centering\small
\begin{tabular}{@{}l>{\raggedright\arraybackslash}p{6.6cm}>{\raggedleft\arraybackslash}p{2cm}@{}}
\toprule
\textbf{Group} & \textbf{Occupations with highest exposure} & \textbf{\% Exposure}\\
\midrule
\textbf{Human} $\pmb{\alpha}$ & Interpreters and Translators  &  76.5 \\
 &  Survey Researchers  &  75.0 \\
 &  Poets, Lyricists and Creative Writers  &  68.8 \\
 &  Animal Scientists  &  66.7 \\
 &  Public Relations Specialists  &  66.7 \\
\midrule
\textbf{Human} $\pmb{\beta}$ & Survey Researchers  &  84.4 \\
 &  Writers and Authors  &  82.5 \\
 &  Interpreters and Translators  &  82.4 \\
 &  Public Relations Specialists  &  80.6 \\
 &  Animal Scientists  &  77.8 \\
\midrule
\textbf{Human} $\pmb{\zeta}$  &  Mathematicians  &  100.0 \\
 &  Tax Preparers  &  100.0 \\
 &  Financial Quantitative Analysts  &  100.0 \\
 &  Writers and Authors  &  100.0 \\
 &  Web and Digital Interface Designers  &  100.0 \\
 & \multicolumn{2}{l}{\textit{\ \ \ \ Humans labeled 15 occupations as "fully exposed."}} \\
\midrule
\textbf{Model} $\pmb{\alpha}$ & Mathematicians  &  100.0 \\
 &  Correspondence Clerks  &  95.2 \\
 &  Blockchain Engineers  &  94.1 \\
 &  Court Reporters and Simultaneous Captioners  &  92.9 \\
 &  Proofreaders and Copy Markers  &  90.9 \\
\midrule
\textbf{Model} $\pmb{\beta}$ &  Mathematicians  &  100.0 \\
 &  Blockchain Engineers  &  97.1 \\
 &  Court Reporters and Simultaneous Captioners  &  96.4 \\
 &  Proofreaders and Copy Markers  &  95.5 \\
 &  Correspondence Clerks  &  95.2 \\
\midrule
\textbf{Model} $\pmb{\zeta}$ &  Accountants and Auditors  &  100.0 \\
 &  News Analysts, Reporters, and Journalists  &  100.0 \\
 &  Legal Secretaries and Administrative Assistants  &  100.0 \\
 &  Clinical Data Managers  &  100.0 \\
 &  Climate Change Policy Analysts  &  100.0 \\
 & \multicolumn{2}{l}{\textit{\ \ \ \ The model labeled 86 occupations as "fully exposed."}} \\
\midrule
\textbf{Highest variance} & Search Marketing Strategists  &  14.5 \\
 &  Graphic Designers  &  13.4 \\
 &  Investment Fund Managers  & 13.0 \\
 &  Financial Managers  &  13.0 \\
 &  Insurance Appraisers, Auto Damage  & 12.6 \\
\bottomrule
\end{tabular}
\caption{Occupations with the highest exposure according to each measurement. The final row lists the occupations with the highest $\sigma^2$ value, indicating that they had the most variability in exposure scores. Exposure percentages indicate the share of an occupation's task that are exposed to GPTs ($\pmb{\alpha}$) or GPT-powered software ($\pmb{\beta}$ and $\pmb{\zeta}$), where exposure is defined as driving a reduction in time it takes to complete the task by at least 50\% (see exposure rubric \ref{exposure_tax}). As such, occupations listed in this table are those where we estimate that GPTs and GPT-powered software are able to save workers a significant amount of time completing a large share of their tasks, but it does not necessarily suggest that their tasks can be fully automated by these technologies. All tasks are assigned equal weight within an occupation.}
\label{tab:occupations-automatability}
\end{table}

%% file: tables/skill_correlation.tex


\begin{table}[h]\centering\scriptsize
\def\sym#1{\ifmmode^{#1}\else\(^{#1}\)\fi}
\begin{tabular}{@{}l>{\raggedright\arraybackslash}p{2cm}>{\raggedleft\arraybackslash}p{2cm}@{}>{\raggedright\arraybackslash}p{2.5cm}}
\toprule
\textbf{Basic Skill} & $\pmb{\alpha}$ \newline (std err) & $\pmb{\beta}$ \newline (std err) & $\pmb{\zeta}$ \newline (std err) \\
\midrule
 & \multicolumn{2}{l}{\textit{\ \ \ \ All skill importance scores are normalized to be between 0 and 1.}} \\
\midrule
\textbf{Constant} & 0.082*** & -0.112*** & 0.300*** \\
 & (0.011) & (0.011) & (0.057)   \\
 \midrule
\textbf{Active Listening} & 0.128** & 0.214*** & 0.449*** \\
 & (0.047) & (0.043) & (0.027) \\
 \midrule
\textbf{Mathematics} & -0.127*** & 0.161*** & 0.787*** \\
 & (0.026)   & (0.021) & (0.049) \\
 \midrule
\textbf{Reading Comprehension} & 0.153*** & 0.470*** & -0.346*** \\
& (0.041) & (0.037) & (0.017) \\
\midrule
\textbf{Science} & -0.114*** & -0.230*** & -0.346*** \\
& (0.014) & (0.012) & (0.017) \\
\midrule
\textbf{Speaking} & -0.028 & 0.133*** & 0.294*** \\
& (0.039) & (0.033) &    (0.042)    \\
\midrule
\textbf{Writing} & 0.368*** & 0.467*** & 0.566*** \\
& (0.042) &    (0.037) & (0.047) \\
\midrule
\textbf{Active Learning} & -0.157*** & -0.065** & 0.028 \\
& (0.027) & (0.024) & (0.032) \\
\midrule
\textbf{Critical Thinking} & -0.264*** & -0.196*** & -0.129** \\
& (0.036) & (0.033) & (0.042) \\
\midrule
\textbf{Learning Strategies} & -0.072* & -0.209*** & -0.346*** \\
& (0.028) & (0.025) & (0.034) \\
\midrule
\textbf{Monitoring} & -0.067** & -0.149*** & -0.232*** \\
& (0.023) & 0.020) & (0.026) \\
\midrule
\textbf{Programming} & 0.637*** & 0.623*** & 0.609*** \\
& (0.030) & (0.022) & (0.024) \\

\bottomrule
\end{tabular}
\caption{Regression of occupation-level, human-annotated exposure to GPTs on skill importance for each skill in the O*NET Basic skills category, plus the programming skill. Descriptions of the skills may be found in Appendix \ref{sec:skills_definitions}. Task ratings within each occupation for exposure have equal weight.}

\label{tab:skills}
\end{table}

%% file: tables/job_zone_summary.tex
\begin{table}[h!]
\centering
\scriptsize
\begin{tabular}{>{\raggedright\arraybackslash}p{.4cm} >{\raggedright\arraybackslash}p{1.7cm} >{\raggedright\arraybackslash}p{2.2cm} >{\raggedright\arraybackslash}p{3cm} | >{\raggedright\arraybackslash}p{1cm} | >{\raggedright\arraybackslash}p{1.2cm} | >{\raggedleft\arraybackslash}p{.39cm} >{\raggedleft\arraybackslash}p{.39cm} | >{\raggedleft\arraybackslash}p{.39cm}>{\raggedleft\arraybackslash}p{.39cm} | >{\raggedleft\arraybackslash}p{.39cm}>{\raggedleft\arraybackslash}p{.39cm}}
\toprule
\textbf{Job Zone} & \textbf{Preparation Required} & \textbf{Education Required} & \textbf{Example Occupations} & \textbf{Median Income} & \textbf{Tot Emp (000s}) & \textbf{H} $\pmb{\alpha}$ & \textbf{M} $\pmb{\alpha}$ & \textbf{H} $\pmb{\beta}$ & \textbf{M} $\pmb{\beta}$ & \textbf{H} \ $\pmb{\zeta}$\  & \textbf{M} $\pmb{\zeta}$ \\
\midrule
1 & None or little (0-3 months) & High school diploma or GED (otional) & Food preparation workers, dishwashers, floor sanders & \$30,230 & 13,100 & 0.03 & 0.04 & 0.06 & 0.06 & 0.09 & 0.08 \\
2 & Some (3-12 months) & High school diploma & Orderlies, customer service representatives, tellers & \$38,215 & 73,962 & 0.07 & 0.12 & 0.16 & 0.20 & 0.24 & 0.27 \\

3 & Medium (1-2 years) & Vocational school, on-the-job training, or associate's degree & Electricians, barbers, medical assistants & \$54,815 & 37,881 & 0.11 & 0.14 & 0.26 & 0.32 & 0.41 & 0.51 \\
4 & Considerable (2-4 years) & Bachelor's degree & Database administrators, graphic designers, cost estimators & \$77,345 &   56,833 & 0.23 & 0.18 & 0.47 & 0.51 & 0.71 & 0.85 \\
5 & Extensive (4+ years) & Master's degree or higher & Pharmacists, lawyers, astronomers & \$81,980 & 21,221 & 0.23 & 0.13 & 0.43 & 0.45 & 0.63 & 0.76 \\
\bottomrule
\end{tabular}
\caption{Mean exposure to GPTs by job zone. For each job zone, we also present the median of median annual income for each constituting occupation in USD, and the total number of workers in all occupations for that job zone, in the thousands. Task weights are equal for all tasks.}
\end{table}

%% file: tables/on_the_job_training.tex
\begin{table}[]\scriptsize
    \centering
\begin{tabular}{l|rr|rr|rr|rr}
\toprule
{\textbf{On The Job Training Required}} & \textbf{Median Income} & \textbf{Tot Emp (000s)} & H  $\pmb{\alpha}$ &  M $\pmb{\alpha}$ &  H $\pmb{\beta}$ &  M $\pmb{\beta}$ &  H $\pmb{\zeta}$ &  M $\pmb{\zeta}$ \\
 &      &      &      &      &      &      \\
\midrule
None                           &  \$77,440    &   90,776  & 0.20 & 0.16 & 0.42 & 0.46 & 0.63 & 0.76 \\
Apprenticeship                 &  \$55,995    &   3,066  & 0.01 & 0.02 & 0.04 & 0.06 & 0.07 & 0.10 \\
Internship/residency          &  \$77,110    &  3,063  & 0.16 & 0.06 & 0.36 & 0.38 & 0.55 & 0.71 \\
Short-term on-the-job training      & \$33,370     & 66,234  & 0.11 & 0.15 & 0.21 & 0.25 & 0.32 & 0.34 \\
Moderate-term on-the-job training     & \$46,880     &  31,285 & 0.09 & 0.12 & 0.21 & 0.25 & 0.32 & 0.38 \\
Long-term on-the-job training      & \$48,925   & 5,070   & 0.08 & 0.10 & 0.18 & 0.22 & 0.28 & 0.33 \\
\bottomrule
\end{tabular}
    \caption{Mean exposure scores for occupations, grouped by level of on-the-job training required to attain competency in the job. Alongside exposure scores, we display the median of median annual income for each occupation, as well as the total number of workers in each group, in thousands. Task weights are equal within an occupation and sum to one.}
    \label{tab:on_the_job_training}
\end{table}

%% file: tables/autoscores_summary.tex
{
\def\sym#1{\ifmmode^{#1}\else\(^{#1}\)\fi}
\begin{tabular}{l*{1}{cccccccc}}
\hline\hline
                    &         Min&  25th Perc.&      Median&   75th Perc&         Max&        Mean&   Std. Dev.&       Count\\
\hline
GPT-4 Exposure Rating 1&        0.00&        0.13&        0.34&        0.50&        1.00&        0.33&        0.22&         750\\
GPT-4 Exposure Rating 2&        0.00&        0.09&        0.24&        0.40&        0.98&        0.26&        0.20&         750\\
Human Exposure Rating&        0.00&        0.09&        0.29&        0.47&        0.84&        0.29&        0.21&         750\\
Software (Webb)     &        1.00&       25.00&       50.00&       75.00&      100.00&       50.69&       30.05&         750\\
Robot (Webb)        &        1.00&       22.00&       52.00&       69.00&      100.00&       48.61&       28.61&         750\\
AI (Webb)           &        1.00&       28.00&       55.00&       82.00&      100.00&       54.53&       29.65&         750\\
Suitability for Machine Learning&        2.60&        2.84&        2.95&        3.12&        3.55&        2.99&        0.18&         750\\
Normalized Routine Cognitive&       -3.05&       -0.46&        0.10&        0.63&        3.42&        0.07&        0.86&         750\\
Normalized Routine Manual&       -1.81&       -0.81&       -0.11&        0.73&        2.96&        0.05&        1.01&         750\\
AI Occupational Exposure Score     &        1.42&        3.09&        3.56&        4.04&        6.54&        3.56&        0.70&         750\\
Frey \& Osborne Automation&        0.00&        0.07&        0.59&        0.88&        0.99&        0.50&        0.38&         681\\
Log Avg. Salary     &       10.13&       10.67&       11.00&       11.34&       12.65&       11.02&        0.45&         749\\
\hline\hline
\end{tabular}
}

%% file: tables/autoscores_regression.tex
{
\def\sym#1{\ifmmode^{#1}\else\(^{#1}\)\fi}
\begin{tabular}{l*{8}{D{.}{.}{-1}}}
\toprule
                    &\multicolumn{2}{c}{GPT-4 Exposure Rating 1}&\multicolumn{2}{c}{GPT-4 Exposure Rating 2}&\multicolumn{2}{c}{Human Exposure Rating}  \\\cmidrule(lr){2-3}\cmidrule(lr){4-5}\cmidrule(lr){6-7}\cmidrule(lr){8-9}
                    &\multicolumn{1}{c}{(1)}         &\multicolumn{1}{c}{(2)}         &\multicolumn{1}{c}{(3)}         &\multicolumn{1}{c}{(4)}                          &\multicolumn{1}{c}{(5)}         &\multicolumn{1}{c}{(6)}         \\
\midrule
Software (Webb)     &     0.00113\sym{***}&     0.00123\sym{***}&     0.00111\sym{***}&     0.00119\sym{***}&     0.00096\sym{***}&     0.00101\sym{***}\\
                    &   (0.00031)         &   (0.00031)         &   (0.00031)         &   (0.00031)         &   (0.00031)         &   (0.00031)         \\
\addlinespace
Robot (Webb)        &    -0.00378\sym{***}&    -0.00405\sym{***}&    -0.00377\sym{***}&    -0.00399\sym{***}&    -0.00371\sym{***}&    -0.00383\sym{***}\\
                    &   (0.00032)         &   (0.00031)         &   (0.00034)         &   (0.00033)         &   (0.00029)         &   (0.00028)         \\
\addlinespace
AI (Webb)           &     0.00080\sym{***}&     0.00090\sym{***}&     0.00036         &     0.00045         &     0.00067\sym{**} &     0.00071\sym{**} \\
                    &   (0.00030)         &   (0.00029)         &   (0.00030)         &   (0.00030)         &   (0.00030)         &   (0.00030)         \\
\addlinespace
Suitability for Machine Learning&     0.29522\sym{***}&     0.26888\sym{***}&     0.28468\sym{***}&     0.26245\sym{***}&     0.19514\sym{***}&     0.18373\sym{***}\\
                    &   (0.04503)         &   (0.04418)         &   (0.04404)         &   (0.04342)         &   (0.03990)         &   (0.03886)         \\
\addlinespace
Normalized Routine Cognitive&     0.06601\sym{***}&     0.06868\sym{***}&     0.04743\sym{***}&     0.05015\sym{***}&     0.03568\sym{***}&     0.03659\sym{***}\\
                    &   (0.00886)         &   (0.00894)         &   (0.00872)         &   (0.00879)         &   (0.00671)         &   (0.00669)         \\
\addlinespace
Normalized Routine Manual&    -0.11147\sym{***}&    -0.11371\sym{***}&    -0.09390\sym{***}&    -0.09561\sym{***}&    -0.11045\sym{***}&    -0.11152\sym{***}\\
                    &   (0.00785)         &   (0.00789)         &   (0.00817)         &   (0.00818)         &   (0.00741)         &   (0.00744)         \\
\addlinespace
AI Occupational Exposure Score     &     0.00993         &     0.02465\sym{**} &    -0.01537         &    -0.00265         &     0.00630         &     0.01252         \\
                    &   (0.01107)         &   (0.01059)         &   (0.01160)         &   (0.01114)         &   (0.00918)         &   (0.00845)         \\
\addlinespace
Frey \& Osborne Automation&    -0.03024\sym{*}  &    -0.03950\sym{**} &    -0.00364         &    -0.01217         &    -0.03890\sym{**} &    -0.04253\sym{**} \\
                    &   (0.01835)         &   (0.01841)         &   (0.02007)         &   (0.01972)         &   (0.01883)         &   (0.01858)         \\
\addlinespace
Log Avg. Salary     &     0.05804\sym{***}&                     &     0.04863\sym{***}&                     &     0.02531         &                     \\
                    &   (0.01870)         &                     &   (0.01860)         &                     &   (0.01727)         &                     \\
\addlinespace
Constant            &    -1.12937\sym{***}&    -0.45743\sym{***}&    -0.96117\sym{***}&    -0.39935\sym{***}&    -0.47078\sym{*}  &    -0.17706         \\
                    &   (0.26859)         &   (0.15327)         &   (0.26365)         &   (0.15017)         &   (0.24684)         &   (0.13256)         \\
\midrule
N                   &   680.00000         &   681.00000         &   680.00000         &   681.00000         &   680.00000         &   681.00000         \\
$R^2$                  &     0.68741         &     0.68212         &     0.60737         &     0.60198         &     0.71213         &     0.71126         \\
\bottomrule
\end{tabular}
}

%% file: exposure_taxonomy.tex
\# E Exposure Taxonomy

Consider the most powerful OpenAI large language model (LLM) This model can complete many tasks that can be formulated as having text input and text output where the context for the input can be captured in 2000 words. The model also cannot draw up-to-date facts (those from <1 year ago) unless they are captured in the input.

Assume you are a worker with an average level of expertise in your role trying to complete the given task. You have access to the LLM as well as any other existing software or computer hardware tools mentioned in the task. You also have access to any commonly available technical tools accessible via a laptop (e.g. a microphone, speakers, etc.). You do not have access to any other physical tools or materials. 

Please label the given task according to the taxonomy below. 

\#\# E0 – No exposure

Label tasks E0 if direct access to the LLM through an interface like ChatGPT or the OpenAI playground cannot reduce the time it takes to complete this task with equivalent quality by half or more.

If a task requires a high degree of human interaction (for example, in person demonstrations) then it should be classified as E0.

\#\# E1 – Direct exposure

Label tasks E1 if direct access to the LLM through an interface like ChatGPT or the OpenAI playground alone can reduce the time it takes to complete the task with equivalent quality by at least half. This includes tasks that can be reduced to:
- Writing and transforming text and code according to complex instructions, 
- Providing edits to existing text or code following specifications,
- Writing code that can help perform a task that used to be done by hand,
- Translating text between languages,
- Summarizing medium-length documents, 
- Providing feedback on documents,
- Answering questions about a document, or 
- Generating questions a user might want to ask about a document.

\#\# E2 – Exposure by LLM-powered applications

Label tasks E2 if having access to the LLM alone may not reduce the time it takes to complete the task by at least half, but it is easy to imagine additional software that could be developed on top of the LLM that would reduce the time it takes to complete the task by half. This software may include capabilities such as:
- Summarizing documents longer than 2000 words and answering questions about those documents
- Retrieving up-to-date facts from the Internet and using those facts in combination with the LLM capabilities
- Searching over an organization’s existing knowledge, data, or documents and retreiving information

Examples of software built on top of the LLM that may help complete worker activities include: 
- Software built for a home goods company that quickly processes and summarizes their up-to-date internal data in customized ways to inform product or marketing decisions 
- Software that is able to suggest live responses for customer service agents speaking to customers in their company’s customer service interface
- Software built for legal purposes that can quickly aggregate and summarize all previous cases in a particular legal area and write legal research memos tailored to the law firm’s needs
- Software specifically designed for teachers that allows them to input a grading rubric and upload the text files of all student essays and have the software output a letter grade for each essay 
- Software that retrieves up-to-date facts from the internet and uses the capabilities of the LLM to output news summaries in different languages

\#\# E3 – Exposure given image capabilities

Suppose you had access to both the LLM and a system that could view, caption, and create images. This system cannot take video media as inputs. This system cannot accurately retrieve very detailed information from image inputs, such as measurements of dimensions within an image. Label tasks as E3 if there is a significant reduction in the time it takes to complete the task given access to a LLM and these image capabilities:
- Reading text from PDFs, 
- Scanning images, or
- Creating or editing digital images according to instructions.

\#\# Annotation examples: 

Occupation: Inspectors, Testers, Sorters, Samplers, and Weighers
Task: Adjust, clean, or repair products or processing equipment to correct defects found during inspections.
Label (E0/E1/E2/E3): E0
Explanation: The model does not have access to any kind of physicality, and more than half of the task (adjusting, cleaning and repairing equipment) described requires hands or other embodiment.

Occupation: Computer and Information Research Scientists
Task: Apply theoretical expertise and innovation to create or apply new technology, such as adapting principles for applying computers to new uses.
Label (E0/E1/E2/E3): E1
Explanation: The model can learn theoretical expertise during training as part of its general knowledge base, and the principles to adapt can be captured in the text input to the model.

Activity: Schedule dining reservations.
Label (E0/E1/E2/E3): E2
Explanation: Automation technology already exists for this (e.g. Resy) and it’s unclear what an LLM offers on top of using that technology (no-diff). That said, you could build something that allows you to ask the LLM to make a reservation on Resy for you. (E3)

Activity: Negotiate purchases or contracts.
Label (E0/E1/E2/E3): E2
Explanation: You could have each party transcribe their point of view and then feed this to an LLM to resolve any disputes (E3). That said, many people would need to buy into using new technological tools to accomplish this (system).

Occupation: Allergists and Immunologists
Task: Prescribe medication such as antihistamines, antibiotics, and nasal, oral, topical, or inhaled glucocorticosteroids.
Label (E0/E1/E2/E3): E2
Explanation: The model can provide guesses for different diagnoses and write prescriptions and case notes. However, it still requires a human in the loop using their judgment and knowledge to make the final decision.

—

%% file: tables/education_table.tex
\scriptsize
\centering
\begin{tabular}{l|rr|rr|rr|rr}
\toprule
{} & Median Income & Emp (000s) &  H $\pmb{\alpha}$ &  M $\pmb{\alpha}$ &  H $\pmb{\beta}$ &  M $\pmb{\beta}$ &  H $\pmb{\zeta}$ &  M $\pmb{\zeta}$ \\
 &      &      &      &      &      &      \\
\midrule
No formal educational credential & \$31,900 & 36,187  & 0.05 & 0.06 & 0.10 & 0.10 & 0.15 & 0.15 \\
High school diploma or equivalent & \$45,470 & 67,033  & 0.09 & 0.13 & 0.20 & 0.25 & 0.31 & 0.37 \\
Postsecondary nondegree award  & \$48,315  & 9,636  & 0.07 & 0.15 & 0.19 & 0.28 & 0.31 & 0.41 \\
Some college, no degree      & \$40,970 & 2,898   & 0.23 & 0.34 & 0.39 & 0.53 & 0.55 & 0.72 \\
Associate's degree          & \$60,360  & 3,537   & 0.12 & 0.14 & 0.31 & 0.36 & 0.49 & 0.59 \\
Bachelor's degree           & \$78,375  & 71,698   & 0.23 & 0.17 & 0.47 & 0.51 & 0.70 & 0.84 \\
Master's degree              & \$79,605  & 3,216  & 0.26 & 0.14 & 0.46 & 0.44 & 0.66 & 0.74 \\
Doctoral or professional degree & \$82,420 & 5,290   & 0.21 & 0.13 & 0.41 & 0.43 & 0.60 & 0.74 \\
\bottomrule
\end{tabular}

%% file: tables/bottom_jobs.tex
\begin{table}[H]
\centering
\begin{tabular}{@{}l@{}}
\toprule
\textbf{Occupations with no labeled exposed tasks} \\
\midrule
Agricultural Equipment Operators \\
Athletes and Sports Competitors \\
Automotive Glass Installers and Repairers \\
Bus and Truck Mechanics and Diesel Engine Specialists \\
Cement Masons and Concrete Finishers \\
Cooks, Short Order \\
Cutters and Trimmers, Hand \\
Derrick Operators, Oil and Gas \\
Dining Room and Cafeteria Attendants and Bartender Helpers \\
Dishwashers \\
Dredge Operators \\
Electrical Power-Line Installers and Repairers \\
Excavating and Loading Machine and Dragline Operators, Surface Mining \\
Floor Layers, Except Carpet, Wood, and Hard Tiles \\
Foundry Mold and Coremakers \\
Helpers--Brickmasons, Blockmasons, Stonemasons, and Tile and Marble Setters \\
Helpers--Carpenters \\
Helpers--Painters, Paperhangers, Plasterers, and Stucco Masons \\
Helpers--Pipelayers, Plumbers, Pipefitters, and Steamfitters \\
Helpers--Roofers \\
Meat, Poultry, and Fish Cutters and Trimmers \\
Motorcycle Mechanics \\
Paving, Surfacing, and Tamping Equipment Operators \\
Pile Driver Operators \\
Pourers and Casters, Metal \\
Rail-Track Laying and Maintenance Equipment Operators \\
Refractory Materials Repairers, Except Brickmasons \\
Roof Bolters, Mining \\
Roustabouts, Oil and Gas \\
Slaughterers and Meat Packers \\
Stonemasons \\
Tapers \\
Tire Repairers and Changers \\
Wellhead Pumpers \\
\bottomrule
\end{tabular}
\caption{All 34 occupations for which none of our measures labeled any tasks as exposed.}
\label{table:top_jobs}
\end{table}